\documentclass[prc,twocolumn,aps,tightenlines,showpacs,floatfix]{revtex4-1}
\usepackage{amsmath}
\usepackage{mwe}
\usepackage[caption=false]{subfig}
\usepackage{cancel}
\usepackage{bm}
\usepackage{color}
\usepackage{graphicx}
\usepackage{amssymb}

\usepackage{multirow}

\usepackage{dcolumn}     
\newcolumntype{.}{D{.}{.}{-1}}

\def\beq{\begin{equation}}
\def\eeq{\end{equation}}
\def\beqy{\begin{eqnarray}}
\def\eeqy{\end{eqnarray}}
\newcommand{\bra}{{\langle}}
\newcommand{\ket}{{\rangle}}

\newcommand{\inelastic}{\omega}
\newcommand{\elastic}{\cancel{\omega}}

\newcommand{\myvec}[1]{{\boldsymbol {#1}  }}

\newcommand{\nn}{\nonumber\\}
\newcommand{\bs}[1]{{\boldsymbol {#1}  }}
\usepackage{booktabs}
\usepackage{siunitx}
\usepackage{tabularx}
\usepackage[colorlinks,breaklinks]{hyperref}
\begin{document}

\title{Probing uncertainties of nuclear structure corrections in light muonic atoms}

\author{O.~J.~Hernandez$^{a,b}$}
\author{C.~Ji$^{c}$}%
\email{jichen@mail.ccnu.edu.cn}
\author{S.~Bacca$^{a}$}
\author{N.~Barnea$^d$}
\affiliation{
$^a$ Institut f\"ur Kernphysik and PRISMA Cluster of Excellence, Johannes Gutenberg-Universit\"at Mainz, 55128 Mainz, Germany \\
$^b$ Department of Physics and Astronomy, University of British Columbia, Vancouver, BC, V6T 1Z4, Canada\\
$^c$ Key Laboratory of Quark and Lepton Physics (MOE) and Institute of Particle Physics,  Central China Normal University, Wuhan 430079,China\\
$^d$ Racah Institute of Physics, The Hebrew University, Jerusalem 91904, Israel\\
}


\begin{abstract}
Recent calculations of  nuclear structure corrections to the Lamb shift in light muonic atoms are based on an expansion in a  parameter $\eta$, where only terms up to second order are retained. The parameter $\eta$ can be shown to be proportional to $\sqrt{m_r/m_p}$, where $m_r$ is the reduced mass of the muon--nucleus system and  $m_p$ is the proton mass, so that it is small and the expansion is expected to converge. However, practical implementations show that the $\eta$ convergence may be slower than expected. In this work we probe the uncertainties due to this expansion using a different formalism, which is based on a multipole expansion of the longitudinal and transverse response functions and was first introduced by Leidemann and Rosenfelder~\cite{Rosenfelder_1983,Leidemann_1995}. We refer to this alternative expansion as the $\eta$-less formalism. We generalize this formalism to account for the cancellation of elastic terms such as the third Zemach moment (or Friar moment) and embed it in a computationally efficient framework. We implement and test this approach in the case of muonic deuterium. The comparison of results in the point nucleon limit for both methods achieve sub-percent agreement. When nucleon form factors are introduced we find a $4\%$ and $2\%$ difference in the third Zemach moment and nuclear polarizability, respectively, compared to the $\eta$-less expansion, indicating that the nucleon form factor approximations in Ref.~\cite{Ji_2018} should be improved. However, we find that the sum of these terms removes this dependence and the uncertainty due to the $\eta$-expansion and the related second-order approximation in the nucleon form factors amounts only to 0.2$\%$ and thus is fully justified in muonic deuterium. This computationally efficient framework paves the way to further studies in light muonic systems with more than two nucleons, where controlling and reducing uncertainties in nuclear structure corrections is key to the experimental efforts of the CREMA collaboration.

\end{abstract}

\pacs{21.10.Ky, 23.20.-g, 23.20.Js, 27.20.+n}

\maketitle


\section{Introduction}
Light muonic atoms have attracted a lot of attention in  recent years.  
The discovery of the proton radius puzzle \cite{Pohl_2010,Antognini_2013} and the subsequent deuteron radius puzzle \cite{Pohl_2016} have driven the experimental effort to probe the Lamb shift in heavier muonic atoms, such as muonic helium, and possibly muonic lithium in the future \cite{Pohl_private}. The Lamb shift $\delta_{\rm LS}$ is related to the charge radius of a nucleus $r_{\rm nucl}$ by
\begin{equation}
\label{eq1}
\delta_{\rm LS} = \delta_{\rm QED}+\mathcal{A}_{\rm OPE}r^2_{\rm nucl}+\delta_{\rm TPE}.
\end{equation}
The term $\delta_{\rm QED}$ denotes the quantum electrodynamics (QED) corrections dominated by the vacuum-polarization and self-energy effects of the muon. The next two terms, $\mathcal{A}_{\rm OPE}r^2_{\rm nucl}$ and $\delta_{\rm TPE}$, are the nuclear structure corrections stemming from one- and two-photon exchange (TPE), respectively. The values of $\delta_{\rm QED}$, $\mathcal{A}_{\rm OPE}$, and $\delta_{\rm TPE}$ in Eq.~(\ref{eq1}) need to be provided by theory and are key to extract the radius $r^2_{\rm nucl}$ from a spectroscopic measurement of the Lamb shift.  For muonic atom experiments, $\delta_{\rm TPE}$  is the bottleneck in the exploitation of the experimental precision and its uncertainty drives the precision by which the radius can be extracted from Eq.~(\ref{eq1}). By convention, $\delta_{\rm TPE}$ is evaluated as a sum of terms that depend on dynamics of the atomic nucleus and the nucleon, denoted with an $A$ and $N$, respectively. These terms are further broken up into the elastic Zemach component ($\delta^{A/N}_{\rm Zem}$) and the inelastic polarizability contribution ($\delta^{A/N}_{\rm pol}$), so that
\begin{equation}
\delta^{A/N}_{\rm TPE} = \delta^{A/N}_{\rm Zem}+\delta^{A/N}_{\rm pol}.\label{eq: nuclear-nucleon two-photon exchange}
\end{equation} 
The elastic contributions are embodied by the third Zemach moment~\cite{Zemach:1956zz} (or Friar moment~\cite{Friar_1977}) and are thus denoted with the label ``${\rm Zem}$''.

For muonic deuterium ($\mu$D), 
the TPE contributions $\delta_{\rm TPE}$ have been calculated by several independent theoretical works~\cite{Pachucki_2011,Friar_2013,Hernandez_2014,Pachucki_2015,Carlson_2014,Hernandez_2018}.
In the $\mu$D experiment, the CREMA collaboration determination of the deuteron charge radius~\cite{Pohl_2016} deviates by 5.6$\sigma$ with respect to the CODATA-2014 evaluation \cite{CODATA_2014}. Another less significant difference in the deuteron was found between the experimentally extracted
muonic isotope-shift radius and the electronic one \mbox{${r_D^2-r_p^2}$}~\cite{Pohl_2016,Jentschura11}. 
A thorough analysis of all systematic and statistical uncertainties in the theoretical calculations of $\delta_{\rm TPE}$ was carried out in Ref.~\cite{Hernandez_2018} and could not account for the deuteron radius puzzle or the smaller isotope-shift disagreement. In that work, the TPE contributions at fifth order in $\alpha$ were calculated which also included the logarithmic Coulomb corrections at order $\alpha^6 \ln\alpha$. The uncertainty arising from other uncalculated $\alpha^6$ diagrams were estimated. These estimates were consistent with the work of Ref.~\cite{Pachucki_2018} where the three-photon exchange contributions were calculated. However, recent work by Kalinowski \cite{Kalinowski_2019} has found that the effect of the vacuum polarization in $\mu$D of order $\alpha^6$ was larger than expected and shifted the central value of $\delta_{\rm TPE}$ towards the experiment. It remains an open question to demonstrate that a consistent calculation of all other relevant contributions at $\alpha^6$ order are as negligible as expected.

In the work of Refs.~\cite{Pachucki_2011,Ji_2013,Friar_2013,Hernandez_2014,Pachucki_2015,Hernandez_2016}
 the nuclear polarizability corrections originating from the TPE were calculated by a multipole expansion of the operator $\eta= \sqrt{2m_r\omega_N} |\myvec{R}-\myvec{R}'|$, where $m_r$ is the reduced lepton-nuclear mass, $\omega_N$ is the nuclear excitation energy and $|\myvec{R}-\myvec{R}'|$ is the ``virtual" distance that a nucleon travels inside the nucleus during the TPE process. The nuclear excitation energy is of the order $\omega_N \sim Q^2/(2 m_p)$ where $m_p$ is the proton mass and $Q$ is the momentum of the moving proton. Based on the uncertainty principle $Q \sim 1/|\myvec{R}-\myvec{R}'|$, one can see that $\eta\sim \sqrt{m_r/m_p}$, and therefore $\eta$ is expected to be small. 
This theoretical formalism refered to as the ``$\eta$-expansion", will be briefly reviewed in the next section. 

The $\eta$-expansion is useful for obtaining closed form expressions that can be easily computed. It can also be used to extract elastic contributions from $\delta^A_{\rm pol}$ that cancel corresponding terms from the elastic TPE. However, for this $\eta$-expansion method to be valid and accurate, the parameter $\eta$ must be much smaller than $1$. For the muonic deuterium and helium-4 the expansion was observed to converge quickly \cite{Hernandez_2016,Hernandez_2014,Ji_2013}. However, for the muonic tritium and helium-3 the $\eta$-expansion was observed to converge slowly \cite{Nevo_Dinur_2016,Javiers_thesis}, leading to a larger estimated truncation uncertainty from this method. 

Extending the $\eta$-expansion to higher orders is technically complicated and motivates us to investigate the more systematic framework introduced by Leidemann and Rosenfelder \cite{Rosenfelder_1983,Leidemann_1995} for calculating the TPE, which will be referred to as the ``$\eta$-less" method in this paper. Leidemann and Rosenfelder's method has not been used in recent calculations of the nuclear TPE due to its higher computational complexity. Furthermore, the original formulation did not allow the extraction of the elastic terms from the polarizability contribution, thus did not exploit direct cancellations of the Zemach moments \cite{Pachucki_2011,Friar_2013}. In this work, we revisit the original Leidemann and Rosenfelder formalism and show that the Lanczos sum rule method \cite{Dinur_2014} can be adapted to tackle these calculations. We also establish how the elastic term implicitly contained in the polarizability can be extracted and removed for any general TPE diagram. 

We apply our new formalism to $\mu$D, and compare it with the $\eta$-expansion method. To this end we utilize two nuclear force models. We first use nuclear potentials from pionless effective field theory ($\cancel{\pi}$EFT) at next-to-next-to-leading order \cite{Friar_2013}. This allows for the derivation of analytic results that are used as a benchmark between $\eta$-less and $\eta$-expansion calculations. Next we use nucleon-nucleon potentials derived from chiral effective field theory ($\chi$EFT) \cite{Entem_2003}.

The paper is organized as follows. Section \ref{Section: Eta expansion} provides a brief pedagogical overview of the $\eta$-expansion and motivates the alternative $\eta$-less formalism for the non-relativistic case, followed by Section \ref{Section: covariant etaless expansion} that introduces the covariant formalism of Ref.~\cite{Rosenfelder_1983,Leidemann_1995} and generalizes the $\eta$-less expansion. In Section \ref{Section: elastic and inelastic cancellations}, we outline how the elastic terms that are implicitly included in the $\eta$-less formalism can be extracted from the inelastic TPE diagram. The latter will partially cancel the contributions from the elastic two-photon diagrams. Finally, in Section \ref{section: results} we present the results of this formalism for muonic deuterium, compare them against the results from $\eta$-expansion and discuss future extensions of the work to other muonic atoms. The details of the $\cancel{\pi}$EFT calculations are given in Appendix \ref{Section: Pionless-EFT at N2LO }, corrections from finite nucleon size effects are explained in Appendix \ref{Section: finite size corrections} and the details of the multipole expansion in $\eta$-less formalism are in Appendix \ref{section: multipole expansion}.

\section{Theoretical framework}\label{section: theoretical framework}

\subsection{The $\eta$-expansion}\label{Section: Eta expansion}

In the work of Refs.~\cite{Pachucki_2011,Ji_2018,Hernandez_2018,Ji_2013,Dinur_2014,Friar_2013} the nuclear polarizability contributions were derived from a power expansion of the small parameter $\eta$. In this section we briefly review this approach.

The non-relativistic limit of the nuclear polarizability $\delta^{\rm NR}_{\rm pol}$ can be calculated from the second order perturbation theory through the matrix element \cite{Ji_2018}
\begin{equation}
\delta^{\rm NR}_{\rm pol} = \langle N_0 \mu | \Delta H G \Delta H   | N_0 \mu \rangle,
\label{eq:def nuclear polarizability correction}
\end{equation}
where $|\mu N_0\rangle = |\mu \rangle \otimes | N_0 \rangle$ is the outer product of the nuclear ground state $|N_0\rangle$ and the muon $2S$-state wave function $|\mu \rangle$. The operator $G$ is the inelastic Green's function of the Hamiltonian, $H= H_{\rm nucl}+H_{\mu}$, where $H_\mu$ is the muon Hamiltonian and $H_{\rm nucl}$ is the nuclear one. The nuclear structure perturbs to the point-Coulomb interaction by
\begin{align}
\Delta H &= \sum_{a}^Z \Delta V(\myvec{r},\myvec{R}_a), 
\end{align}
where
\begin{align}
\Delta V(\myvec{r},\myvec{R}_a) &= -\alpha\left( \frac{1}{|\myvec{r}-\myvec{R}_a|} - \frac{1}{r} \right).
\end{align}
Here $\myvec{r}$ is the position of the muon relative to the center of the nucleus, $\myvec{R}_a$ denotes the coordinates of the $a$-th proton relative to the nuclear center, $\alpha$ is the fine structure constant and $Z$ is the charge number of the nucleus. Integrating over $\myvec{r}$ allows Eq.~\eqref{eq:def nuclear polarizability correction} to be re-written in terms of the muon matrix element $W$ as
\begin{equation}
\delta^{\rm NR}_{\rm pol} = \sum_{N\neq N_0} \int d^3R \ d^3R' \rho^p_N(\myvec{R})W(\myvec{R},\myvec{R}',\omega_N)\rho^p_N(\myvec{R}'), \label{eq: delta_A_pol with W}
\end{equation}
where $\rho^{p}_N(\myvec{R})$ is the nuclear point-proton transition density, which is defined by
\begin{equation}
\label{eq:rhop_N}
\rho_N^{p}(\myvec{R}) = \bra N | \frac{1}{Z}\sum_{a}^{Z} \delta(\myvec{R}-\myvec{R}_a)|N_0\ket.
\end{equation}

The muon matrix element can be written as
\begin{align}
W(\myvec{R},\myvec{R}',\omega_N) =& -Z^2 |\phi_\mu(0)|^2 \int \frac{d^3q}{(2\pi)^3} \left(\frac{4\pi \alpha}{q^2} \right)^2 
\nn
 &\times \left(1-e^{i\myvec{q}\cdot \myvec{R}} \right) \frac{1}{\frac{q^2}{2m_r}+\omega_N}\left(1-e^{-i\myvec{q}\cdot \myvec{R}'} \right), \label{eq: W in terms of exp}
\end{align}
with $|\phi_\mu(0)|^2=(m_r Z\alpha)^3/8\pi$ denoting the norm of the muon $2S$ wave function.
After carrying out the integral over the momentum $\myvec{q}$ of the virtual photon, $W$ becomes a function of the dimensionless parameter $\eta = \sqrt{2m_r\omega_N} |\myvec{R}-\myvec{R}'|$
\begin{align}
W(\myvec{R},\myvec{R}',\omega_N) =& -\frac{\pi}{m^2_r} (Z \alpha)^2 |\phi_\mu(0)|^2 \left( \frac{2m_r}{\omega_N}\right)^{3/2}
\nn
&\times \frac{1}{\eta}\left( e^{-\eta} -1 + \eta - \frac{1}{2}\eta^2 \right).
\end{align}
As discussed in the introduction, this dimensionless parameter has been qualitatively argued to be of the order $\sqrt{\frac{m_\mu}{m_p}} < 1$ allowing $W$ to be expanded in powers of $\eta$. 

This $\eta$-expansion allows the non-relativistic expression to be a sum of leading, subleading, etc., contributions with respect to the associated powers of $\eta$
\begin{equation}
\delta^{\rm NR}_{\rm pol} = \delta^{(0)}_{\rm NR}+\delta^{(1)}_{\rm NR}+\delta^{(2)}_{\rm NR} + ..., 
\end{equation}
where $\delta^{(0)}_{\rm NR}$ is dominated by the dipole correction $\delta^{(0)}_{D1}$ \cite{Ji_2018}, and the sub-leading term $\delta^{(1)}_{\rm NR}$ is the sum of the elastic contributions $\delta^{(1)}_{R3}$ and $\delta^{(1)}_{Z3}$ \cite{Ji_2018} with
\begin{align}
\delta^{(1)}_{R3} =&- \frac{\pi }{3} m_r (Z\alpha)^2 |\phi_\mu(0)|^2 \notag\\
 &\times\iint d^3R d^3R' |\myvec{R}-\myvec{R}'|^3 \rho^{(pp)}_0(\myvec{R},\myvec{R}'), \label{eq: Definition of R3} 
\end{align}
and
\begin{align}
\delta^{(1)}_{Z3} =& \frac{\pi }{3} m_r (Z\alpha)^2 |\phi_\mu(0)|^2 \notag\\ 
&\times \iint d^3R d^3R' |\myvec{R}-\myvec{R}'|^3 \rho^p_0(\myvec{R})\rho^p_0(\myvec{R}'), \label{eq: Definition of Z3}
\end{align}
where $\rho^{p}_0$ is the ground state point-proton density and $\rho^{(pp)}_0$ denotes the proton-proton correlation density. It is important to note that the latter term exactly cancels out the elastic contribution $\delta^A_{\rm Zem}$, i.e., $\delta^A_{Z3}=-\delta^A_{\rm Zem}$ in Eq.~\eqref{eq: nuclear-nucleon two-photon exchange}. In the work of \cite{Pachucki_2011,Ji_2013,Friar_2013,Hernandez_2014,Pachucki_2015,Nevo_Dinur_2016,Ji_2018}, the $\eta$-expansion was carried out up to sub-sub-leading order. Higher order terms in this expansion lead to non-analytic expressions that are difficult to calculate and have until now only been estimated \cite{Pachucki_2011,Ji_2018}. 

The $\eta$-expansion can be circumvented by introducing the multipole expansion of $\exp(i \bs{q}\cdot \bs{R})$ into Eq.~\eqref{eq: W in terms of exp}, integrating over the angles $\hat{q}$ and plugging the result back into Eq.~\eqref{eq: delta_A_pol with W}. This results in the TPE correction
\begin{equation}
\delta^{\rm NR}_{\rm pol} = -8 (Z\alpha)^2 |\phi_\mu(0)|^2 
\int\limits_0^\infty dq \int\limits_{\omega_{\rm th}}^{\infty} d\omega \ K_{\rm NR}(q,\omega)S_L(q,\omega), \label{eq: Non-relativistic two-photon exchange}
\end{equation}
where the non-relativistic Kernel is defined as
\begin{equation}
K_{\rm NR}(q,\omega) = \frac{1}{q^2(\frac{q^2}{2m_r}+\omega)}, \label{eq: Definition of the Non-relativisitic Kernel} 
\end{equation}
and $S_L$ is the longitudinal nuclear response function defined in the next section. The expression in Eq.~\eqref{eq: Non-relativistic two-photon exchange} provides a more systematic method to calculate the nuclear structure corrections, however, it is not apparent that it contains the elastic term $\delta^{(1)}_{Z3}$ that cancels the corresponding term from the elastic diagram. Nonetheless, as we shall demonstrate in Section \ref{Section: elastic and inelastic cancellations} the extraction of the elastic components of a general TPE diagram can be readily accomplished by generalizing the formalism of Refs.~\cite{Rosenfelder_1983,Leidemann_1995} discussed in the next section.

\subsection{$\eta$-less formalism}\label{Section: covariant etaless expansion}

\begin{figure}[h]
\includegraphics[scale=.27]{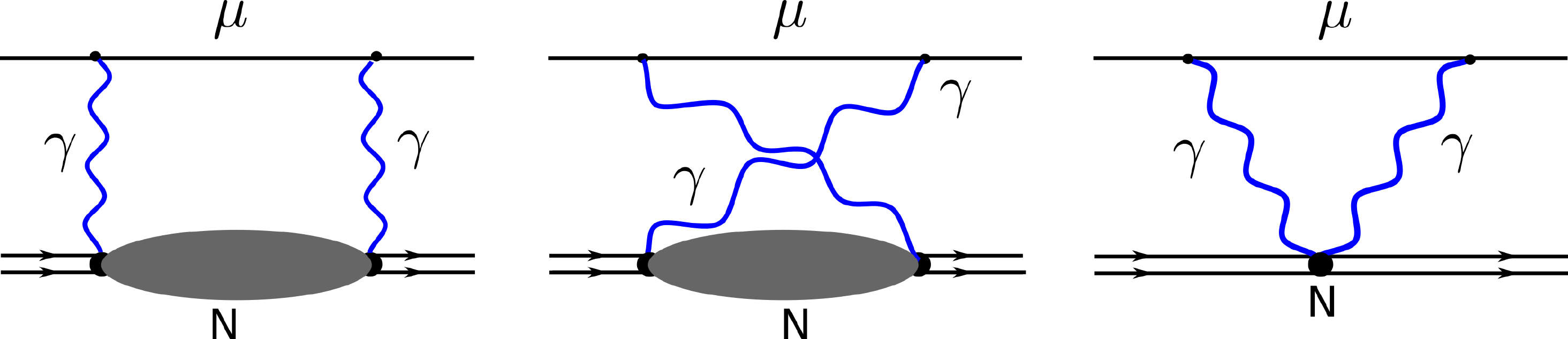}
\par\medskip
\caption{Two-photon exchange diagrams: direct, crossed and seagull diagrams. The grey blob represents the excited states of the nucleus.}
\label{fig:all two-photon exchange diagrams}
\end{figure}

Following the work in Refs.~\cite{Rosenfelder_1983,Leidemann_1995}, the contributions of the TPE diagrams given by Fig.~\ref{fig:all two-photon exchange diagrams} are
\begin{align}
\delta^A_{\rm pol} =& - 8 (Z\alpha)^2 |\phi_\mu(0)|^2 \ 
\text{Im} \int \frac{d^4p}{(2\pi)^4}
D^{\mu \rho}(p)D^{\nu \tau}(-p)
\nn
&\times t_{\mu \nu}(p,k)T_{\rho \tau}(p,-p),
\end{align}
where $D^{\mu \nu}(p)$ is the photon propagator, $t_{\mu \nu}(p,k)$ is the lepton tensor with $k=(m_r,\myvec{0})$ and $T_{\mu \nu}(p,-p)$ is the hadronic tensor. The evaluation of this amplitude in the Coulomb gauge yields
\begin{align}
\delta^A_{\rm pol} =& -8 (Z\alpha)^2 |\phi_\mu(0)|^2 \int\limits_0^\infty dq \int\limits_{\omega_{\rm th}}^{\infty} d\omega 
\left[ K_L(q,\omega)S_L(q,\omega) \right.
\nn
&\left. +K_T(q,\omega)S_T(q,\omega)+K_S(q,\omega)S_T(0,\omega) \right], \label{eq: Total two photon exchange correction}
\end{align}
here we integrate over the nuclear excitation energy $\omega$ and the magnitude of the three-momentum vector $q=|\myvec{q}|$. The lower limit $\omega_{\rm th}$ in the integral is the nuclear threshold energy. $S_L(q,\omega)$ and $S_T(q,\omega)$ are the longitudinal and transverse response functions, respectively. They are defined as
\begin{align}
S_{L}(q,\omega) &= \sum\limits_{N\neq N_0}|\langle N |\tilde{\rho}(q)| N_0 \rangle|^2 \delta(\omega_N - \omega), \label{eq:Definition of Longitudinal response functions} \\
S_T(q,\omega) &=  \sum_{\lambda= \pm 1}\sum\limits_{N\neq N_0}|\langle N |\hat{e}^\dagger_\lambda\cdot \tilde{\myvec{J}}(q)| N_0 \rangle|^2\delta(\omega_N - \omega),
\label{eq:Definition of Transverse response functions}
\end{align}
where $| N_0 \rangle$ and $| N \rangle$ are the ground and excited states of the nucleus, with energies $E_0$ and $E_N$, respectively. $\omega_N= E_N-E_0$ is the nuclear excitation energy. Here the summation over excited states is averaged over their angular momentum projections. The vectors $\hat{e}^\dagger_\lambda$ are the circular transverse polarization vectors. The operators $\tilde{\rho}$ and $\tilde{\myvec{J}}$ are the Fourier transforms of the nuclear charge and current densities, respectively. As in Refs.~\cite{Rosenfelder_1983,Leidemann_1995} the kernels in the integrals of Eq.~\eqref{eq: Total two photon exchange correction} are
\begin{align}
K_{L}(q,\omega) =& \frac{1}{2 E_q}\left[\frac{1}{(E_q -m_r)(\omega+E_q-m_r)} \right. \notag\\
&\left. - \frac{1}{(E_q +m_r)(\omega+E_q+m_r)} \right], \label{eq: Definition of Longitudinal Kernel} \\
K_{T}(q,\omega) =& \frac{q^2}{4m_r^2}K_L(q,\omega) - \frac{1}{4m_r q} \frac{\omega+2q}{(\omega+q)^2}, \label{eq: Definition of Transverse Kernel} \\
K_S(q,\omega) =& \frac{1}{4m_r \omega}\left[\frac{1}{q} - \frac{1}{E_q} \right], \label{eq: Definition of the Seagul Kernel}
\end{align}
where $E_q = \sqrt{m^2_r+q^2}$ is the relativistic energy of the muon. 
It is convenient to separate the transverse response function into the electric $S^{\rm E}_T$ and magnetic $S^{\rm M}_T$ response functions. The nuclear polarization from the TPE is then decomposed into the sum
\begin{align}
\delta^A_{\rm pol} =& \Delta_{L} + \Delta_{T,\rm E}+ \Delta_{T,\rm M},\label{eq: The relation of delta_pol to L TE and TM corrections}
\end{align}
where the subscripts $L$, ($T,{\rm E}$) and ($T,{\rm M}$) denote the longitudinal, transverse-electric and transverse-magnetic corrections, respectively. These corrections are given explicitly by
\begin{align}
  \Delta_{L}  =& - 8 (Z\alpha)^2 |\phi_\mu(0)|^2 
   \int\limits_{0}^{\infty} dq \int\limits_{\omega_{\rm th}}^{\infty} d\omega \ K_{L}(q,\omega) S_{L}(q,\omega), \label{eq: Longitudinal Correction} \\
    \Delta_{T,{\rm E}}  =& - 8(Z\alpha)^2 |\phi_\mu(0)|^2 
    \int\limits_{0}^{\infty} dq \int\limits_{\omega_{\rm th}}^{\infty} d\omega  
    \nn
    &\times \left[ K_{T}(q,\omega)S^{\rm E}_{T}(q,\omega)  
    +K_S(q,\omega)S^{\rm E}_T(0,\omega)  \right], \label{eq: Transverse Electric Correction}\\
  \Delta_{T,{\rm M}}  =& - 8(Z\alpha)^2 |\phi_\mu(0)|^2  
          \int\limits_{0}^{\infty} dq \int\limits_{\omega_{\rm th}}^{\infty}d\omega \  K_{T}(q,\omega) S^{\rm M}_{T}(q,\omega). \label{eq: Transverse Magnetic Correction}
\end{align}
In the non-relativistic limit, $q \ll m_r$, the kernels reduce to their non-relativistic forms
\begin{align}
K_L(q,\omega) &\rightarrow K_{\rm NR}(q,\omega),\\
K_T(q,\omega) &\rightarrow 0,\\
K_S(q,\omega) &\rightarrow 0,
\end{align}
and Eq.~\eqref{eq: Longitudinal Correction} reduces to the expression given by Eq.~\eqref{eq: Non-relativistic two-photon exchange}. Eqs.~(\ref{eq: The relation of delta_pol to L TE and TM corrections}-\ref{eq: Transverse Magnetic Correction}) summarize the calculation of $\delta^A_{\rm pol}$ within the methodology of the $\eta$-less expansion.

\subsection{Subtraction of the elastic part}\label{Section: elastic and inelastic cancellations}

The inelastic TPE calculation in the $\eta$-less expansion implicitly contains contributions from terms related to elastic corrections, such as the Zemach moments that cancel out corresponding terms in the elastic TPE diagram. Here we refer to the elastic TPE diagrams as those where the nucleus remains in the ground state throughout the process, while the inelastic diagrams are those in which there is sufficient energy transfer from the virtual photons to excite or breakup the nucleus. A consistent TPE calculation requires the treatment of both terms. 

Here we describe how these elastic terms can be extracted directly from the $\eta$-less formalism and give the conditions under which there will be a cancellation between the elastic and inelastic contributions for a TPE diagram at order $\alpha^5$ and higher.
\begin{figure}[h]
\begin{minipage}{.5\linewidth}
\subfloat[]{\label{fig:elastic two-photon exchange}\includegraphics[scale=.12]{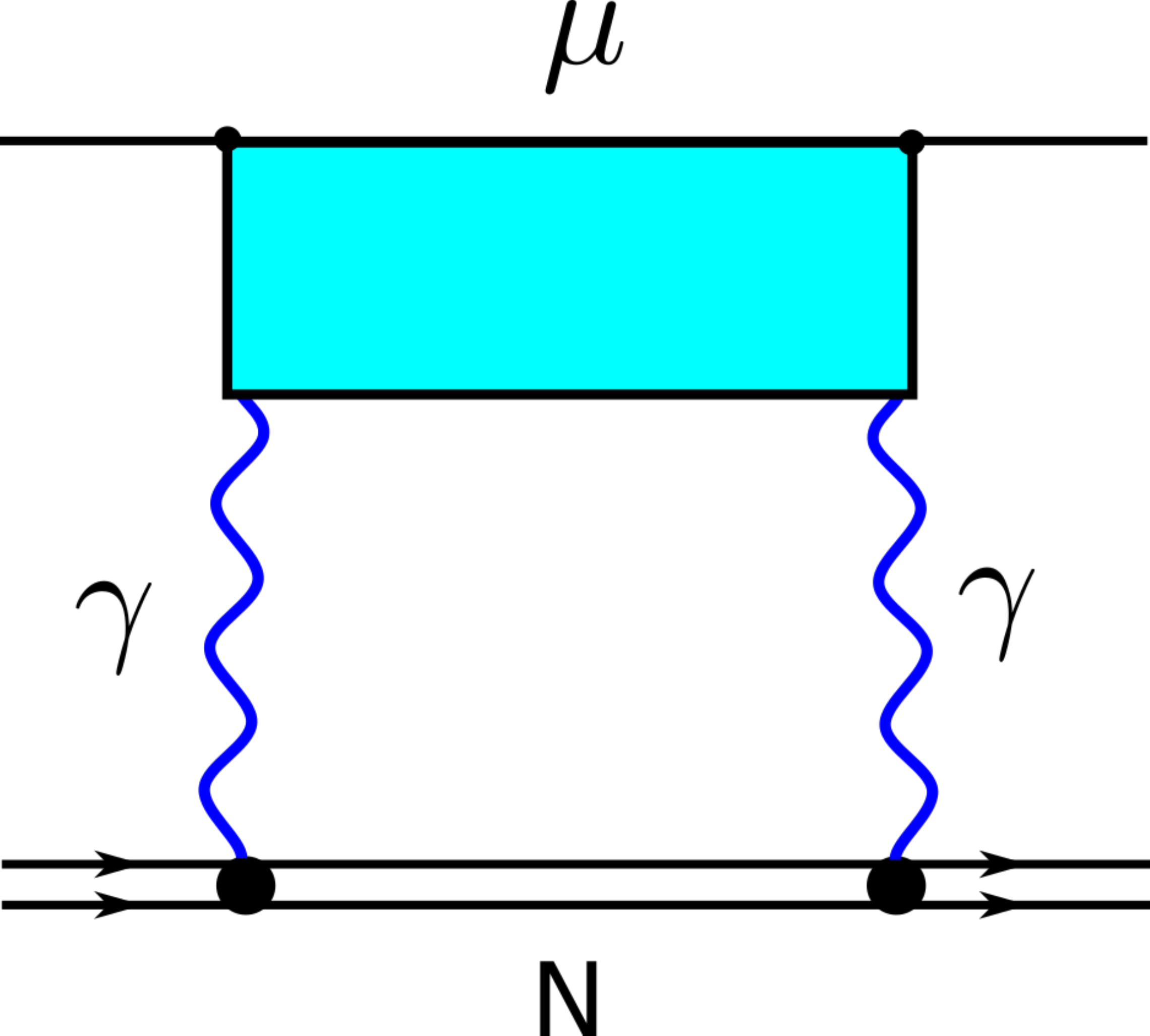}}
\end{minipage}%
\begin{minipage}{.5\linewidth}
\centering
\subfloat[]{\label{fig:inelastic two-photon exchange}\includegraphics[scale=.12]{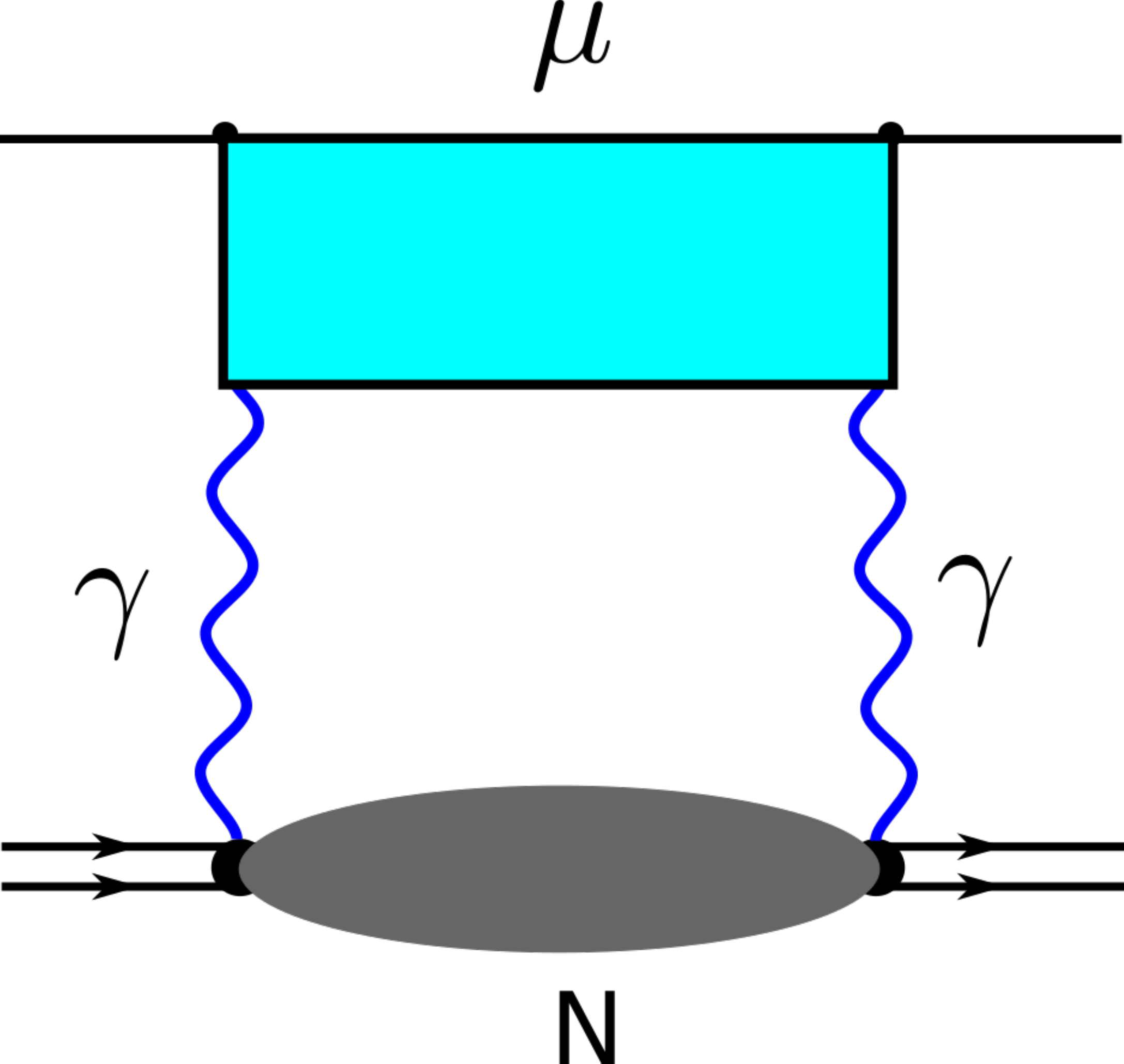}}
\end{minipage}\par\medskip
\caption{Two-photon exchange diagrams: (a) elastic vs (b) inelastic. The light blue square represents electromagnetic processes involving the upper half of the diagram. In the inelastic diagram (b) the excited states of the nucleus are represented by the grey blob.}
\label{fig:main two-photon exchange theorem}
\end{figure}

Fig.~\ref{fig:main two-photon exchange theorem} illustrates the most general elastic and inelastic TPE diagrams, with the light blue square indicating processes that can be inserted into the photon or lepton propagators in the upper half of the diagrams. The contribution of the elastic diagram in Fig.~\ref{fig:elastic two-photon exchange} will be denoted as $\delta^{A}_{\rm el}$, while the inelastic diagram in Fig.~\ref{fig:inelastic two-photon exchange} is $\delta^{A}_{\rm pol}$. The contribution from the elastic diagram is
\begin{align}
\delta^{A}_{\rm el} =& - 8 (Z\alpha)^2 |\phi_\mu(0)|^2 \int\limits_0^\infty dq\ K_{\mu \nu}(q,i0^+) 
\nn
&\times \langle N_0 |\tilde{J}^{\mu}(q)| N_0\rangle \langle N_0 | \tilde{J}^{\nu}(-q)| N_0 \rangle, \label{eq: General Elastic TPE}
\end{align}
where $0^+$ indicates that the $\omega \rightarrow 0^+$ limit is taken. The kernel $K_{\mu \nu}(q,i0^+)$ is determined by the processes considered in the squares of Fig.~\ref{fig:main two-photon exchange theorem}. The matrix elements $\langle N_0 |J^{\mu}(q)| N_0\rangle$ represent the elastic electromagnetic vertices of the nucleus in the ground state. The generalized form of Eq.~\eqref{eq: Total two photon exchange correction} for the inelastic process is
\begin{align}
\delta^{A}_{\rm pol} =  - 8 (Z\alpha)^2 |\phi_\mu(0)|^2  
&\int\limits_0^\infty dq \int\limits_{\omega_{\rm th}}^{\infty} d\omega  \ K_{\mu \nu}(q,\omega) S^{\mu \nu}(q,\omega). \label{eq: General Inelastic TPE}
\end{align}
The general nuclear response function $S^{\mu \nu}(q,\omega)$ is
\begin{align}
S^{\mu \nu}(q,\omega) =&  \sum\limits_{N\neq N_0} \langle  N_0 |\tilde{J}^{\mu}(q) |N \rangle\langle  N |\tilde{J}^{\nu}(-q) |N_0 \rangle 
\nn
&\times \delta(\omega_N-\omega).\label{eq: general response function}
\end{align}
The expression in Eq.~\eqref{eq: general response function} contains the electromagnetic matrix elements connecting the ground state $|N_0\rangle$ to the nuclear excited states $| N \rangle$. The analytical terms that cancel corresponding elastic contributions can be extracted from $\delta^{A}_{\rm pol}$ by separating it into the purely $\omega$-dependent ($\Delta^{\rm inel}_{\inelastic}$) terms and $\omega$-independent ($\Delta^{\rm inel}_{\elastic}$) components as
\begin{equation}
\delta^{A}_{\rm pol}  =   \Delta^{\rm inel}_{\inelastic} +  \Delta^{\rm inel}_{\elastic}.  
\end{equation}
These $\omega$-dependent/independent corrections are given by
\begin{align}
 \Delta^{\rm inel}_{\inelastic} =&  - 8 (Z\alpha)^2 |\phi_\mu(0)|^2 
\int\limits_{0}^{\infty} dq \int\limits_{\omega_{\rm th}}^{\infty} d\omega 
\nn
&\times \left[ K_{\mu\nu}(q,\omega)- K_{\mu\nu}(q,i0^+) \right] 
S^{\mu \nu}(q,\omega), \label{eq: omega dependent correction} 
\\
\Delta^{\rm inel}_{\elastic}  =&  - 8 (Z\alpha)^2 |\phi_\mu(0)|^2 
 \int\limits_{0}^{\infty} dq \int\limits_{\omega_{\rm th}}^{\infty}  d\omega 
 \nn 
 &\times K_{\mu\nu}(q,i0^+)S^{\mu  \nu}(q,\omega),\label{eq: omega independent correction}
\end{align}
respectively. For the $\omega$-independent terms, the kernel $K_{\mu \nu}(q,i0^+)$ can be taken out from the integration over $\omega$, so that the remaining integral uses the identity
\begin{align}
 \int\limits_{\omega_{\rm th}}^{\infty} d\omega \ S^{\mu \nu}(q,\omega) =& \langle N_0 |\tilde{J}^{\mu}(q)\tilde{J}^{\nu}(-q)| N_0 \rangle \nn
   -& \langle N_0 |\tilde{J}^{\mu}(q)| N_0\rangle \langle N_0 | \tilde{J}^{\nu}(-q)| N_0 \rangle ,
 \label{eq: Integration over Weightless Response Function}
\end{align}
where $\langle N_0 |\tilde{J}^{\mu}(q)\tilde{J}^{\nu}(-q)| N_0 \rangle$ is the correlation function of the electromagnetic operators and $\langle N_0 |\tilde{J}^{\mu}(q)| N_0 \rangle$ is directly related to the elastic electromagnetic form factors of the nucleus. 
Therefore, $\Delta^{\rm inel}_{\elastic}$ is divided into two terms
\begin{align}
\Delta^{\rm inel}_{\elastic}  = \Delta_{\rm corr}-\delta^A_{\rm el},
\label{eq: Final}
\end{align}
where the first term is the correlation term defined by
\begin{align}
 \Delta_{\rm corr} =& - 8 (Z\alpha)^2 |\phi_\mu(0)|^2 
 \nn
 \times& \int\limits_0^\infty dq \ K_{\mu\nu}(q,i 0^{+}) \langle N_0 |\tilde{J}^{\mu}(q)\tilde{J}^{\nu}(-q)| N_0 \rangle.
\end{align}
The second term in Eq.~\eqref{eq: Final} cancels out exactly Eq.~\eqref{eq: General Elastic TPE}. 
Therefore, when contributions of elastic and inelastic TPE diagrams are added, only the correlation and the $\omega$-dependent ($\Delta^{\rm inel}_{\inelastic}$) contributions remain. One indeed obtains   
\begin{equation}
\delta^A_{\rm pol}+\delta^A_{\rm el} = \Delta_{\rm corr}  +  \Delta^{\rm inel}_{\inelastic}.\label{eq: inelastic diagram decomposition}
\end{equation}

For the contributions to Lamb shift at order $\alpha^5$, $\delta_{\rm el}^A$ in Eq.~\eqref{eq: General Elastic TPE} can reduce to the Zemach term $\delta_{\rm Zem}$. This reduction is very general and applies to transitions other than the Lamb shift such as the hyperfine splitting~\cite{Friar:2005je,Friar:2005yv}. Here we only demonstrate it in the point-nucleon limit using the $\eta$-less formalism. The nuclear response functions, which contribute to the Lamb shift, are the longitudinal and transverse ones defined in Eqs.~(\ref{eq:Definition of Longitudinal response functions}, \ref{eq:Definition of Transverse response functions}). When taking the limit $\omega \rightarrow i0^+$, the transverse kernel vanishes, and the longitudinal covariant kernel reduces to the non-relativistic one:
\begin{align}
K_T(q,i0^+)=& 0,\\
K_L(q,i0^+)=& K_{\rm NR} (q,i0^+) = \frac{2m_r}{(q^2+i0^+)^2}.
\end{align}
Therefore, the only remaining elastic TPE component is 
\begin{align}
\delta_{\rm el}^A 
=& -16 m_r (Z\alpha)^2 |\phi_\mu(0)|^2 
  \mathcal{P}\int\limits_0^\infty  \frac{dq}{q^4} |\langle N_0 | \tilde{\rho}^p(q)| N_0  \rangle|^2  
  \nn
=& - \delta^{(1)}_{Z3} = \delta^{(1)}_{\rm Zem},
\label{eq: Z3 correlation integral}
\end{align}
where  $\delta^{(1)}_{Z3}$, derived in the $\eta$-expansion formalism in Eq.~\eqref{eq: Definition of Z3}, cancels exactly the Zemach term in the point-nucleon limit, i.e., $\delta^{(1)}_{\rm Zem}$. Similarly, the correlation term yields
\begin{align}
\Delta_{\rm corr} = &
-16 m_r (Z\alpha)^2 |\phi_\mu(0)|^2
\nn
&\times \mathcal{P} \int\limits_0^\infty   \frac{dq}{q^4} \langle N_0 |  \tilde{\rho}^p(q)\tilde{\rho}^{p,\dagger}(q) | N_0  \rangle 
\nn
=& \delta^{(1)}_{R3},
\label{eq: R3 correlation integral}
\end{align}
where $\delta^{(1)}_{R3}$, given in Eq.~\eqref{eq: Definition of R3}, is the proton-proton correlation term from the $\eta$-expansion. Combining with Eq.~\eqref{eq: Final}, the $\omega$-independent nuclear polarizability correction leads to 
\begin{equation}
\Delta^{\rm inel}_{\elastic} = \delta^{(1)}_{R3} - \delta^{(1)}_{\rm Zem}.  
\end{equation}

Therefore, the polarizability contribution in the $\eta$-less expression in Eq.~\eqref{eq: inelastic diagram decomposition} can be written as 
\begin{equation}
\delta^A_{\rm pol} = \Delta^{\rm inel}_{\inelastic}+ \delta^{(1)}_{R3}-\delta^{(1)}_{\rm Zem},  
\end{equation}
where $\Delta^{\rm inel}_{\inelastic}$ are the $\omega$-dependent nuclear polarizability corrections. This example establishes how the elastic contributions can be extracted from the $\eta$-less expansion formalism. For muonic deuterium or tritium, the correlation term $\delta^{(1)}_{R3}=0$ since the nucleus has only one proton.

\section{Numerical Procedures}\label{section: numerical procedures}

In this section we outline the numerical methods used to calculate the nuclear structure corrections in light muonic atoms within the previously outlined framework. The tools needed are the multipole decomposition for the charge and current density operators, followed by the use of the Lanczos sum rule (LSR) method that makes the required calculations amenable to computation.

Using the multipole expansion from Appendix \ref{section: multipole expansion} on the longitudinal and transverse response functions the nuclear structure corrections in Eqs.~(\ref{eq: Longitudinal Correction}, \ref{eq: Transverse Electric Correction}, \ref{eq: Transverse Magnetic Correction}) are calculated as a sum of terms with different photon multipolarities 
\begin{equation}
\Delta_{x} = \sum_{\mathcal{J}=0}\Delta_{\mathcal{J},x}.
\end{equation}
Here the symbol $x$ denotes $L,(T,{\rm E})$ or $(T,{\rm M})$. In the non-relativisitic limit, $\Delta_L \rightarrow \delta^{\rm NR}_{\rm pol}$ and $\Delta_{T,{\rm E/M}} \rightarrow 0$. In practice the integrals in Eqs.~\eqref{eq: Longitudinal Correction}-\eqref{eq: Transverse Magnetic Correction} are generalized sum rules of the nuclear response function, Eq.~\eqref{eq: general response function}, given by
\begin{align}
I = \int\limits_0^\infty dq \int\limits_{\omega_{\rm th}}^{\infty} d\omega \ K_{\mu \nu}(q,\omega)S^{\mu \nu}(q,\omega),\label{eq: GSR equation}
\end{align}
with an arbitrary kernel function $K_{\mu \nu}(q,\omega)$. The calculation of this integral requires the diagonalization of the complete Hamiltonian which is computationally impractical. To render this problem tractable one may use the LSR method \cite{Dinur_2014} to carry out the integral over the nuclear excitation energies $\omega$, followed by a Gaussian quadrature discretization to integrate over $q$. The LSR method allows for the efficient computation of sum rules by approximating the full $M \times M$ Hamiltonian matrix with a tridiagonal matrix $H_{M'}$ of dimension $M'$ using the recursive Krylov subspace, where $M'$ is determined by the number of Lanczos iterations. In this Lanczos basis, Eq.~\eqref{eq: GSR equation} becomes  
\begin{align}
I_{M'} =& \sum_i^{N_q} W_i \langle N_0|\tilde{J}^{\dagger \mu}(q_i) \tilde{J}^{\nu}(q_i)|N_0 \rangle 
\nn
&\times \sum_{m\neq 0}^{M'}|U_{m0}|^2 K_{\mu \nu}(q_i ,\omega_m). \label{eq: Lanczos discretization}
\end{align}
Here the integral over the nuclear excitation energy $\omega$ in Eq.~\eqref{eq: GSR equation} is carried out as a sum rule by integrating over all of the discretized energy states $\omega_m \equiv E_{m} -E_{0}$ where $E_{m}$ is the $m$-th eigenvalue of $H_{M'}$ and $U$ is the unitary transformation matrix that diagonalizes $H_{M'}$. The points $q_i$ and $W_i$ for $i=1,\ldots N_q$ are the Gaussian quadrature grid points and weights, respectively, for the momentum integral in Eq.~\eqref{eq: GSR equation}. Within the LSR framework, the low-lying eigenstates and spectral moments converge after a relatively small number of Lanczos iterations $M'$, where $M'$ is typically much smaller than $M$, i.e., the dimension of the original Hamiltonian, allowing the generalized sum rules to be calculated very efficiently.

\subsection{Implementation}

For the deuteron case, the model space is small and it is possible to use the full diagonalization variant of the Hamiltonian when evaluating the sum rules. The two-body system was solved using the truncated harmonic oscillator basis expansion \cite{Javiers_thesis}. The integrals over the momentum were carried out over a discrete Gaussian quadrature grid of 100 points up to a maximum $q$ estimated by the ultraviolet cut-off of the oscillator basis \mbox{$\Lambda_{\rm UV} \approx \sqrt{( 2 N_{\rm Max}+ 7) m_N \Omega/\hbar}$}, where $\Omega$ is the oscillator frequency and $N_{\rm Max}$ is the model space size~\cite{Binder_2016}. This is much less computationally expensive than calculating the full response function on a grid of $\omega$ and $q$ as in Refs.~\cite{Rosenfelder_1983,Leidemann_1995}. For nuclei heavier than the deuteron, the Lanczos variant of Eq.~\eqref{eq: Lanczos discretization} can be used.

\section{Results}\label{section: results}

In this section we provide the results of the $\eta$-less formalism for muonic deuterium 
using both $\cancel{\pi}$EFT and $\chi$EFT interactions. The former is calculated analytically and the latter is evaluated numerically.
In the text that follows, the calculated values are quoted to three significant digits. For the analytical $\cancel{\pi}$EFT case, the results in Table \ref{table: tabulated non-relativistic results pionless-EFT} are quoted up to four digits, while for the results with the $\chi$EFT nuclear forces in Tables \ref{table: tabulated non-relativistic results}-\ref{table: tabulated relativistic and FF results} we quote the values up to their estimated numerical uncertainty. The sources of numerical uncertainty considered for the $\chi$EFT force calculations are the oscillator frequency $\hbar \Omega$, the model space size $N_{\rm Max}$, the maximum momentum value $q_{\rm max}$ used in the integration of the responses and the number of quadrature points $N_q$ used in the momentum integrals. For each $\mathcal{J}$, the total numerical uncertainty is the quadrature sum of these contributions. These numerical uncertainties are the only ones estimated in Tables \ref{table: tabulated non-relativistic results}-\ref{table: tabulated relativistic and FF results}.

\subsection{Analytical $\cancel{\pi}$ EFT}

Before tackling the muonic deuterium calculation with a $\chi$EFT potential, we adapt the formalism of $\cancel{\pi}$EFT at next-to-next-to-leading order to compute the nuclear structure TPE corrections in $\mu$D in the non-relativistic, point-nucleon limit as in Eq.~\eqref{eq: Non-relativistic two-photon exchange}.

\begin{table}[h]
\centering
\begin{tabular}{l|ll} \hline\
{$\mathcal{J}$} & {$\Delta_{L,\mathcal{J}}$}  &  {$\delta^{\rm NR}_{\rm pol}$ }  \\ \hline \vspace{-2mm}
 &                               &                          \\
0 &-5.559$\times 10^{-2}$ &  -5.559$\times 10^{-2}$ \\
1 &  -1.451 &  -1.5066 \\
2 &   -6.455$\times 10^{-2}$ &  -1.5711 \\
3 &   -1.182$\times 10^{-2}$ &  -1.5830 \\
4 &   -3.723$\times 10^{-3}$ &  -1.5867 \\
5 &   -1.545$\times 10^{-3}$ &  -1.5882 \\
6 &   -7.571$\times 10^{-4}$ &  -1.5890 \\
7 &   -4.145$\times 10^{-4}$ &  -1.5894 \\
8 &   -2.457$\times 10^{-4}$ &  -1.5896 \\
9 &   -1.546$\times 10^{-4}$ &  -1.5898 \\
10 &  -1.019$\times 10^{-4}$ &  -1.5899 \\
11 &  -6.972$\times 10^{-5}$ &  -1.5900 \\
12 &  -4.918$\times 10^{-5}$ &  -1.5900 \\
13 &  -3.56$\times 10^{-5}$ &   -1.5901 \\
14 &  -2.631$\times 10^{-5}$ &  -1.5901 \\
15 &  -1.981$\times 10^{-5}$ &  -1.5901 \\
16 &  -1.514$\times 10^{-5}$ &  -1.5901 \\
17 &  -1.174$\times 10^{-5}$ &  -1.5901 \\
18 &  -9.207$\times 10^{-6}$ &  -1.5901 \\
19 &  -7.281$\times 10^{-6}$ &  -1.5901 \\
20 &  -5.822$\times 10^{-6}$ &  -1.5902 \\
\hline
\end{tabular}
\caption{The calculated values of the terms $\Delta_{L,\mathcal{J}}$ that contribute to the two-photon exchange in meV at each multipole $\mathcal{J}$ for $\mu$D with $\cancel{\pi}$EFT at next-to-next-to-leading order, and their running sum $\delta^{\rm NR}_{\rm pol} = \sum_{K=0}^{\mathcal{J}} \Delta_{L,K}$.}
\label{table: tabulated non-relativistic results pionless-EFT}
\end{table}

The formulas to carry out this calculation are presented in Appendix~\ref{Section: Pionless-EFT at N2LO }. In Table \ref{table: tabulated non-relativistic results pionless-EFT}, 
the individual contributions of the term $\Delta_{L,\mathcal{J}}$ for $\mathcal{J}=0,...,20$ are given along with the running sum $\delta^{\rm NR}_{\rm pol} = \sum_{\mathcal{J}=0}^{\mathcal{J}_{\rm max}} \Delta_{L,\mathcal{J}}$, that quickly saturates to -1.5902 meV. In the $\eta$-formalism with $\cancel{\pi}$EFT,  $\delta^{\rm NR}_{\rm pol} = \delta^{(0)}_{D1}+\delta^{(1)}_{Z3}+\delta^{(2)}_{R2}+\delta^{(2)}_{Q}+\delta^{(2)}_{D1D3}$ = -1.590 meV in the same non-relativistic and point-nucleon limit. These results agree within 0.01$\%$ indicating an excellent agreement between the $\eta$-less and $\eta$-expansion methods.

\subsection{$\chi$EFT Case}

Now we consider the nucleon-nucleon potential derived from $\chi$EFT at next-to-next-to-next-to-leading order \cite{Entem_2003}. Calculations are performed as in Ref.~\cite{Hernandez_2018}. We begin with the non-relativistic formalism in Eq.~\eqref{eq: Non-relativistic two-photon exchange} to compute $\delta^{\rm NR}_{\rm pol}$. In Table \ref{table: tabulated non-relativistic results}, each contribution $\Delta_{L,\mathcal{J}}$ is listed from $\mathcal{J}=0,...,20$ along with the running sum.

\begin{table}[h]
\centering
\begin{tabular}{l|ll} \hline\
{$\mathcal{J}$} & {$\Delta_{L,\mathcal{J}}$}  &  {$\delta^{\rm NR}_{\rm pol}$ }  \\ \hline \vspace{-2mm}
 &                               &                          \\
0 & -6.85620(1)$\times 10^{-2}$ & -6.85620(1)$\times 10^{-2}$ \\
1 &   -1.436198(1) & -1.504760(1)  \\
2 & -6.442519(2)$\times 10^{-2}$ & -1.569185(1) \\
3 & -1.18696(1)$\times 10^{-2}$ & -1.581055(1) \\
4 & -3.7455(2)$\times 10^{-3}$ & -1.584800(1)  \\
5 & -1.5570(2)$\times 10^{-3}$ &  -1.586357(1) \\
6 & -7.649(2)$\times 10^{-4}$ &   -1.587122(1) \\
7 & -4.201(3)$\times 10^{-4}$ &   -1.587542(1)  \\
8 & -2.502(3)$\times 10^{-4}$ &   -1.587793(1)  \\
9 & -1.583(4)$\times 10^{-4}$ &   -1.587951(1)  \\
10 & -1.051(4)$\times 10^{-4}$ &  -1.588056(1)  \\
11 & -7.25(4)$\times 10^{-5}$ &  -1.588128(1)  \\
12 & -5.16(4)$\times 10^{-5}$ &   -1.588180(1) \\
13 & -3.77(4)$\times 10^{-5}$ &   -1.588218(1)  \\
14 & -2.82(4)$\times 10^{-5}$ &   -1.588246(1)  \\
15 & -2.15(4)$\times 10^{-5}$ &   -1.588267(2) \\
16 & -1.66(4)$\times 10^{-5}$ & -1.588284(2) \\
17 & -1.31(4)$\times 10^{-5}$ & -1.588297(2) \\
18 & -1.04(3)$\times 10^{-5}$ & -1.588308(2) \\
19 & -8.4(3)$\times 10^{-6}$ & -1.588316(2) \\
20 & -6.8(3)$\times 10^{-6}$ & -1.588323(2) \\
\hline
\end{tabular}
\caption{The calculated values of the terms $\Delta_{L,\mathcal{J}}$ that contribute to the two-photon exchange in meV as a function of the multipole $\mathcal{J}$ using the $\chi$EFT potential for $\mu$D, and their running sum $\delta^{\rm NR}_{\rm pol} = \sum_{K=0}^{\mathcal{J}} \Delta_{L,K}$.}
\label{table: tabulated non-relativistic results}
\end{table}

From Table \ref{table: tabulated non-relativistic results}, we have $\delta^{\rm NR}_{\rm pol}$ = -1.588 meV. The equivalent result in the $\eta$-formalism is $\delta^{\rm NR}_{\rm pol} = \delta^{(0)}_{D1}+\delta^{(1)}_{Z3}+\delta^{(2)}_{R2}+\delta^{(2)}_{Q}+\delta^{(2)}_{D1D3}$ = -1.590 meV. The difference between these values is 0.002 meV indicating excellent agreement between both methods for $\mu$D. Higher order multipole corrections within the $\eta$-expansion formalism for muonic deuterium are thus negligible. Furthermore, we note that only the first few terms $\mathcal{J}=0,1,2,3$ are needed to reach sub-percentage agreement with the final $\delta^{\rm NR}_{\rm pol}$ value.

Next, we consider the evaluation of the relativistic expressions in Eqs.~(\ref{eq: Longitudinal Correction}-\ref{eq: Transverse Magnetic Correction}) in the point-nucleon limits. The individual contributions to the TPE are given in Table \ref{table: tabulated relativistic results} with the final result denoted as $\delta^{\rm R}_{\rm pol}$.

\begin{table*}[ht]
\centering
\begin{tabular}{l|llll} \hline\
{$\mathcal{J}$} & {\quad\quad$\Delta_{L,\mathcal{J}}$} &  {$\Delta_{T,{\rm E},\mathcal{J}}$} & {$\Delta_{T,{\rm M},\mathcal{J}}$}  & {$\delta^{\rm R}_{\rm pol}$} \\ \hline
0 & -6.71358(1)$\times 10^{-2}$ & 0.0 & 0.0 & -6.71358(1)$\times 10^{-2}$  \\
1 & -1.415390(1) & -1.2570(7)$\times 10^{-2}$ & 2.8243(2)$\times 10^{-3}$ &  -1.492271(6)  \\
2 & -6.186081(7)$\times 10^{-2}$ & 1.01894(8)$\times 10^{-4}$ & 5.197(3)$\times 10^{-4}$ &  -1.553511(6)\\
3 & -1.113425(2)$\times 10^{-2}$ & 9.00(1)$\times 10^{-6}$ & 7.05(3)$\times 10^{-5}$ & -1.564565(6)   \\
4 & -3.44382(3)$\times 10^{-3}$ & 2.62(1)$\times 10^{-6}$ & 1.007(6)$\times 10^{-4}$ &  -1.567906(7)  \\
5 & -1.40260(4)$\times 10^{-3}$ & 1.34(1)$\times 10^{-6}$ & 2.40(4)$\times 10^{-5}$ & -1.569283(7) \\
6 & -6.7570(4)$\times 10^{-4}$ & 8.3(2)$\times 10^{-7}$ & 4.9(1)$\times 10^{-5}$ &   -1.569909(7)  \\
7 & -3.6394(5)$\times 10^{-4}$ & 5.7(2)$\times 10^{-7}$ & 1.30(6)$\times 10^{-5}$ & -1.570259(7)  \\
8 & -2.1261(6)$\times 10^{-4}$ & 4.2(2)$\times 10^{-7}$ & 3.0(1)$\times 10^{-5}$ & -1.570442(7)\\
9 & -1.3203(7)$\times 10^{-4}$ & 3.3(2)$\times 10^{-7}$ & 8.2(6)$\times 10^{-6}$ & -1.570565(7) \\
10 & -8.605(8)$\times 10^{-5}$& 2.6(2)$\times 10^{-7}$ & 2.0(2)$\times 10^{-5}$ & -1.570631(7)  \\
11 & -5.830(9)$\times 10^{-5}$ & 2.1(2)$\times 10^{-7}$ & 5.5(6)$\times 10^{-6}$ &  -1.570684(7) \\
12 & -4.08(1)$\times 10^{-5}$ & 1.7(2)$\times 10^{-7}$ & 1.4(2)$\times 10^{-5}$ & -1.570710(7)\\
13 & -2.93(1)$\times 10^{-5}$ & 1.4(2)$\times 10^{-7}$ & 3.9(6)$\times 10^{-6}$ &  -1.570736(7) \\
14 & -2.16(1)$\times 10^{-5}$ & 1.2(2)$\times 10^{-7}$ & 1.0(2)$\times 10^{-5}$ & -1.570747(7) \\
15 & -1.62(1)$\times 10^{-5}$ & 1.0(2)$\times 10^{-7}$ & 2.9(5)$\times 10^{-6}$ &   -1.570760(7)\\
16 & -1.24(1)$\times 10^{-5}$ & 9(2)$\times 10^{-8}$ & 7(2)$\times 10^{-6}$ &  -1.570765(8)  \\
17 & -9.6(1)$\times 10^{-6}$ & 7(2)$\times 10^{-8}$ & 2.2(4)$\times 10^{-6}$ & -1.570773(8)\\
18 & -7.5(1)$\times 10^{-6}$ & 6(2)$\times 10^{-8}$ & 6(1)$\times 10^{-6}$ &  -1.570775(8) \\
19 & -6.0(1)$\times 10^{-6}$ & 5(2)$\times 10^{-8}$ & 1.7(3)$\times 10^{-6}$ &  -1.570779(8)\\
20 & -4.8(1)$\times 10^{-6}$ & 5(2)$\times 10^{-8}$ & 4(1)$\times 10^{-6}$ & -1.570779(8)\\
\hline
\end{tabular}
\caption{The calculated values of the terms that contribute to $\delta^{\rm R}_{\rm pol}$ in meV as a function of the multipole $\mathcal{J}$ for the $\chi$EFT potential. $\delta^{\rm R}_{\rm pol} = \sum_{K=0}^{\mathcal{J}} (\Delta_{L,K}+\Delta_{T,{\rm E},K}+\Delta_{T,{\rm M},K})$ is presented by the running sum.}
\label{table: tabulated relativistic results}
\end{table*}
In the relativistic and point-nucleon limit, the $\eta$-expansion yields  $\delta^{\rm R}_{\rm pol} = \delta^{(0)}_{D1}+\delta^{(0)}_{L}+\delta^{(0)}_{T}+\delta^{(1)}_{Z3}+\delta^{(2)}_{R2}+\delta^{(2)}_{Q}+\delta^{(2)}_{D1D3}$ = -1.573 meV can be compared to the $\eta$-less result $\delta^{\rm R}_{\rm pol} = \Delta_L +\Delta_{T,{\rm E}} + \Delta_{T,{\rm M}} $= -1.571 meV, and amounts to a difference of $0.1 \%$. In the low-$q$ limit the term $\Delta_{T,{\rm E},\mathcal{J}=1}$ = -1.257$\times 10^{-2}$ meV corresponds to the relativistic electric transverse polarization $\delta^{(0)}_{T}$ = -1.248$\times 10^{-2}$ meV from the $\eta$-expansion \cite{Ji_2018} and are in close agreement. The relativistic longitudinal polarizability correction in the $\eta$-expansion, $\delta^{(0)}_{L}$ = 2.913$\times 10^{-2}$  meV, is formally related to the difference between the relativistic and non-relativistic results $\Delta_{L,\mathcal{J}=1}$ in Tables \ref{table: tabulated relativistic results} and \ref{table: tabulated non-relativistic results} amounting to 2.081$\times 10^{-2}$ meV. The difference between these results is accounted for by the fact that in the $\eta$-expansion formalism, the response function is approximated in the low-$q$ limit, while here we have treated the response functions with the full $q$ dependence.\\

We proceed to use the relativistic kernels with nucleon form factors. The parametrizations of the electric and magnetic nucleon elastic form factors are 
\begin{align}
G^p_E(q^2) &= \frac{1}{\left(1 + q^2/\beta^2 \right)^2}, \label{eq: Proton electric FF} \\
G^n_E(q^2) &= \frac{\lambda q^2}{\left(1 + q^2/\beta^2 \right)^3}, \label{eq: Neutron electric FF} 
\end{align}
and
\begin{align}
G^{p/n}_M(q^2) &= \mu_N g_{p/n} G^p_E(q^2),
\end{align}
where $\beta^2 = 12/r^2_p$, $\lambda = -r^2_n/6$, $r^2_n =$ -0.1161(22) fm$^2$ and $r_p = 0.84087(39)~\textrm{fm}$ are parameters given in Refs.~\cite{Ji_2018, Perdrisat_2007}. The magnetic $g$ factors of the proton and neutron are $g_p$=5.585694702(17) and $g_n$=-3.82608545(90), respectively, while $\mu_N$ is the nuclear magneton. 

In the $\eta$-formalism of Ref.~\cite{Ji_2018}, nucleon size effects were included by approximating the nucleon form factors in Eq.~\eqref{eq: Proton electric FF} and \eqref{eq: Neutron electric FF} to linear order in $q^2$. This approximation is valid in the $\eta$-expansion framework, but diverges in the $\eta$-less formalism. Therefore, we carried out the $\eta$-less calculations using the full form factors. This result including relativistic effects and nucleon form factors is denoted as $\delta^A_{\rm pol}$  and are given in Table \ref{table: tabulated relativistic and FF results}.

\begin{table*}[ht]
\centering
\begin{tabular}{l|llll} \hline\
{$\mathcal{J}$} & {\quad\quad$\Delta_{L,\mathcal{J}}$} &  {$\Delta_{T,{\rm E},\mathcal{J}}$} & {$\Delta_{T,{\rm M},\mathcal{J}}$}  & {$\delta^A_{\rm pol}$} \\ \hline
0 & -6.27430(1)$\times 10^{-2}$ & 0.0 & 0.0 & -6.27430(1)$\times 10^{-2}$ \\
1 &-1.389312(1)  & -1.2729(8)$\times 10^{-2}$ &  2.7400(2) $\times 10^{-3}$ & -1.462044(7)  \\
2 & -5.687366(7)$\times 10^{-2}$ & 9.42909(5)$\times 10^{-5}$ & 3.7810338(1)$\times 10^{-4}$ &  -1.518446(7)  \\
3 & -8.66690(1)$\times 10^{-3}$ & 5.6589911(7)$\times 10^{-6}$ & 3.4725465(8)$\times 10^{-5}$  & -1.527072(7)  \\
4 & -2.499426(5)$\times 10^{-3}$ &  1.189995(1)$\times 10^{-6}$  & 3.046054(2)$\times 10^{-5}$ & -1.529540(7)  \\
5 & -7.91031(2)$\times 10^{-4}$ & 3.379354(4)$\times 10^{-7}$  & 5.54470(1)$\times 10^{-6}$   & -1.530325(7)\\
6 & -3.73996(1)$\times 10^{-4}$ & 1.83995(1)$\times 10^{-7}$ &  7.40018(2)$\times 10^{-6}$ &  -1.530692(7) \\
7 & -1.467700(5) $\times 10^{-4}$ & 7.10098(2)$\times 10^{-8}$ & 1.67017(1)$\times 10^{-6}$ & -1.530837(7)  \\
8 &  -8.89794(3)$\times 10^{-5}$ & 5.39528(9)$\times 10^{-8}$ & 2.56595(3)$\times 10^{-6}$ & -1.530923(7)  \\
9 & -3.81065(2)$\times 10^{-5}$ & 2.25638(2)$\times 10^{-8}$ & 6.4689(1)$\times 10^{-7}$ & -1.530960(7) \\
10 & -2.72129(1)$\times 10^{-5}$ & 2.05594(9)$\times 10^{-8}$ & 1.06086(4)$\times 10^{-6}$ &  -1.530986(7) \\
11 & -1.211088(8)$\times 10^{-5}$ & 8.7809(2)$\times 10^{-9}$ & 2.8741(1)$\times 10^{-7}$ &  -1.530998(7)  \\
12 & -9.77783(7) $\times 10^{-6}$ & 9.0617(9)$\times 10^{-9}$ & 4.8931(4)$\times 10^{-7}$ & -1.531008(7) \\
13 & -4.41778(4)$\times 10^{-6}$ & 3.8693(2)$\times 10^{-9}$ & 1.3972(1)$\times 10^{-7}$ & -1.531012(7) \\
14 & -3.93714(4)$\times 10^{-6}$ & 4.3939(7)$\times 10^{-9}$ & 2.4379(3)$\times 10^{-7}$ & -1.531016(7) \\
15 &  -1.78341(2)$\times 10^{-6}$ & 1.8585(2)$\times 10^{-9}$ & 7.2510(8)$\times 10^{-8}$ &  -1.531017(7) \\
16 & -1.72722(2)$\times 10^{-6}$  & 2.2826(5)$\times 10^{-9}$ & 1.2879(3)$\times 10^{-7}$ &  -1.531019(7) \\
17 & -7.7872(1)$\times 10^{-7}$ &  9.523(1)$\times 10^{-10}$  & 3.9588(6)$\times 10^{-8}$ & -1.531020(7)  \\
18 &  -8.1058(1)$\times 10^{-7}$ & 1.2503(4)$\times 10^{-9}$ & 7.129(2)$\times 10^{-8}$ & -1.531020(7) \\
19 & -3.62163(8)$\times 10^{-7}$ & 5.134(1)$\times 10^{-10}$ & 2.2520(5)$\times 10^{-8}$ & -1.531021(7) \\
20 & -4.01814(8)$\times 10^{-7}$ & 7.145(2)$\times 10^{-10}$ & 4.101(2)$\times 10^{-8}$  & -1.531021(7)  \\
\hline
\end{tabular}
\caption{The calculated values of the terms that contribute to $\delta^A_{\rm pol}$ in meV as a function of the multipole $\mathcal{J}$ for the $\chi$EFT potential.  $\delta^A_{\rm pol} = \sum_{K=0}^{\mathcal{J}} (\Delta_{L,K}+\Delta_{T,{\rm E},K}+\Delta_{T,{\rm M},K})$ is presented by the running sum.}
\label{table: tabulated relativistic and FF results}
\end{table*}

In Table \ref{table: tabulated relativistic and FF results} the inclusion of nucleon form factors reduces the magnitude of the individual contributions to $\delta^A_{\rm pol}$ with respect to Table \ref{table: tabulated relativistic results} in the point-nucleon limit. This is expected because the nucleon form factors suppress the strength of the response functions at high $q$-values. The total sum of all contributions in Table \ref{table: tabulated relativistic and FF results} is $\delta^A_{\rm pol}$= -1.531 meV. 

\begin{figure}[h]
\includegraphics[scale=.23]{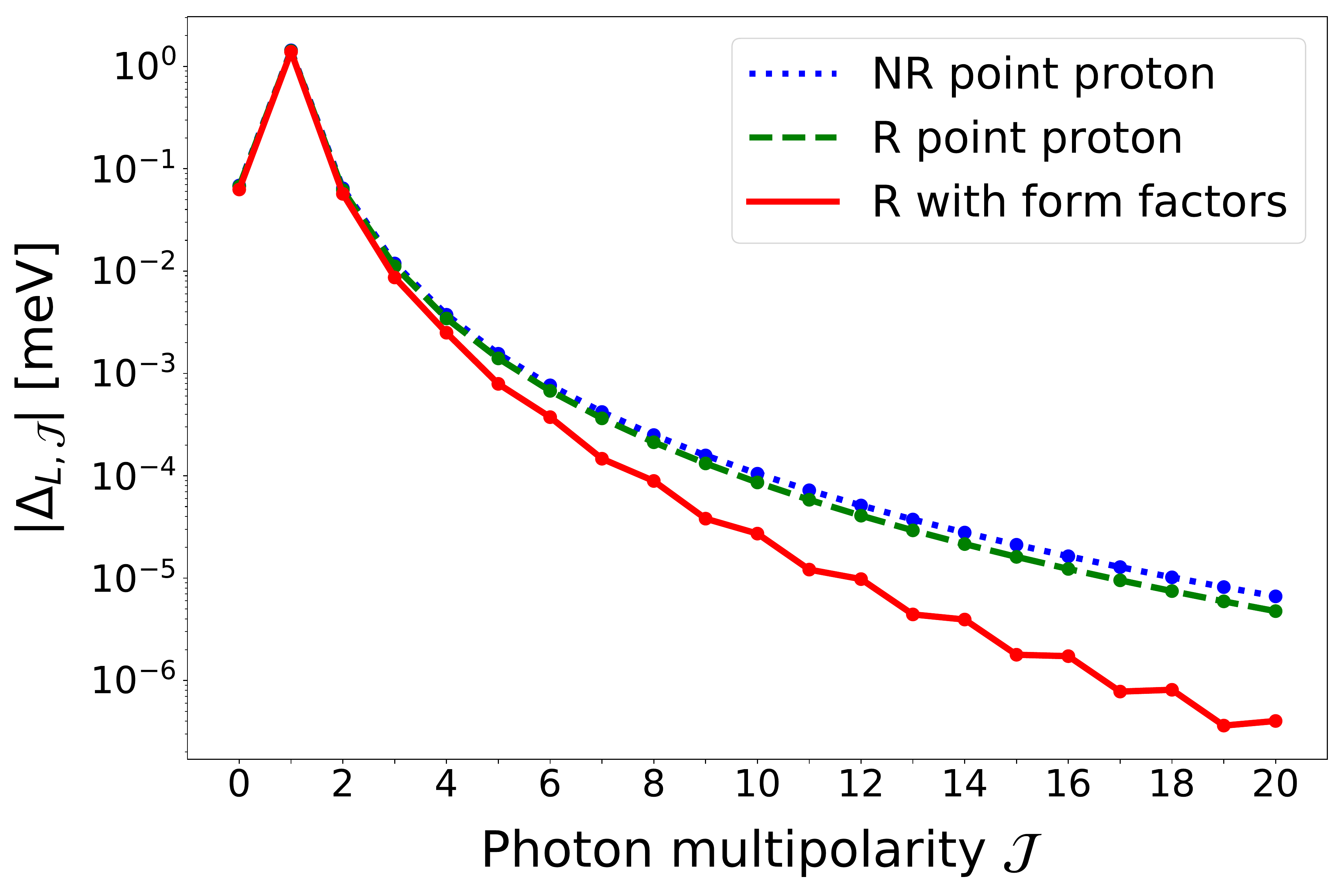}
\caption{The absolute value of the longitudinal corrections in the non-relativisitic (NR) point-nucleon limit, in the relativistic (Rel) point-nucleon limit, and with nucleon form factors as a function of the multipolarity $\mathcal{J}$.}
\label{fig: Convergence of Delta_L}
\end{figure}

\begin{figure}[h]
\subfloat{
	\label{fig: Convergence of Delta_T_el}
	\includegraphics[scale=0.22]{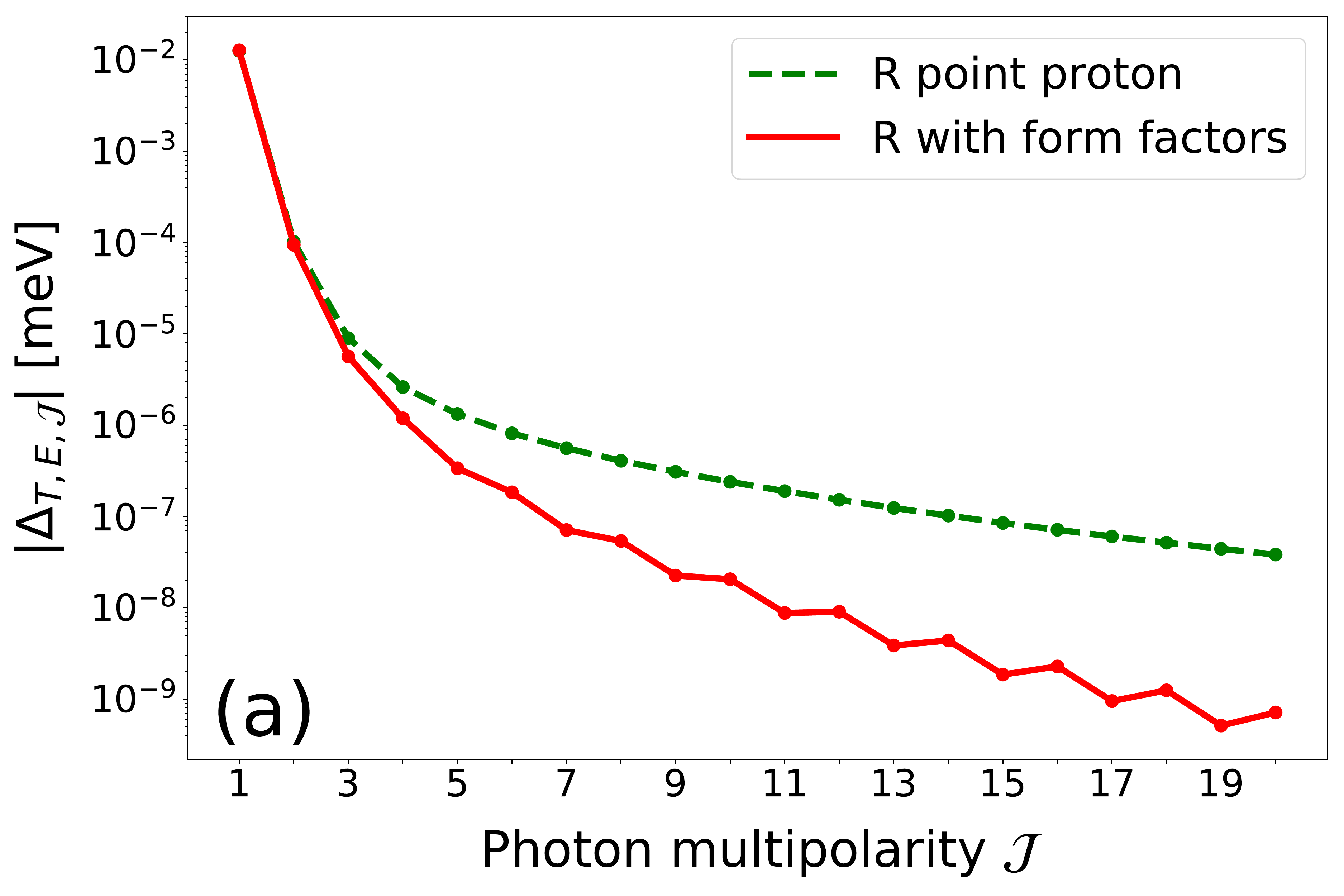} 
	} 
 
\subfloat{
	\label{fig: Convergence of Delta_T_mag}
	\includegraphics[scale=0.22]{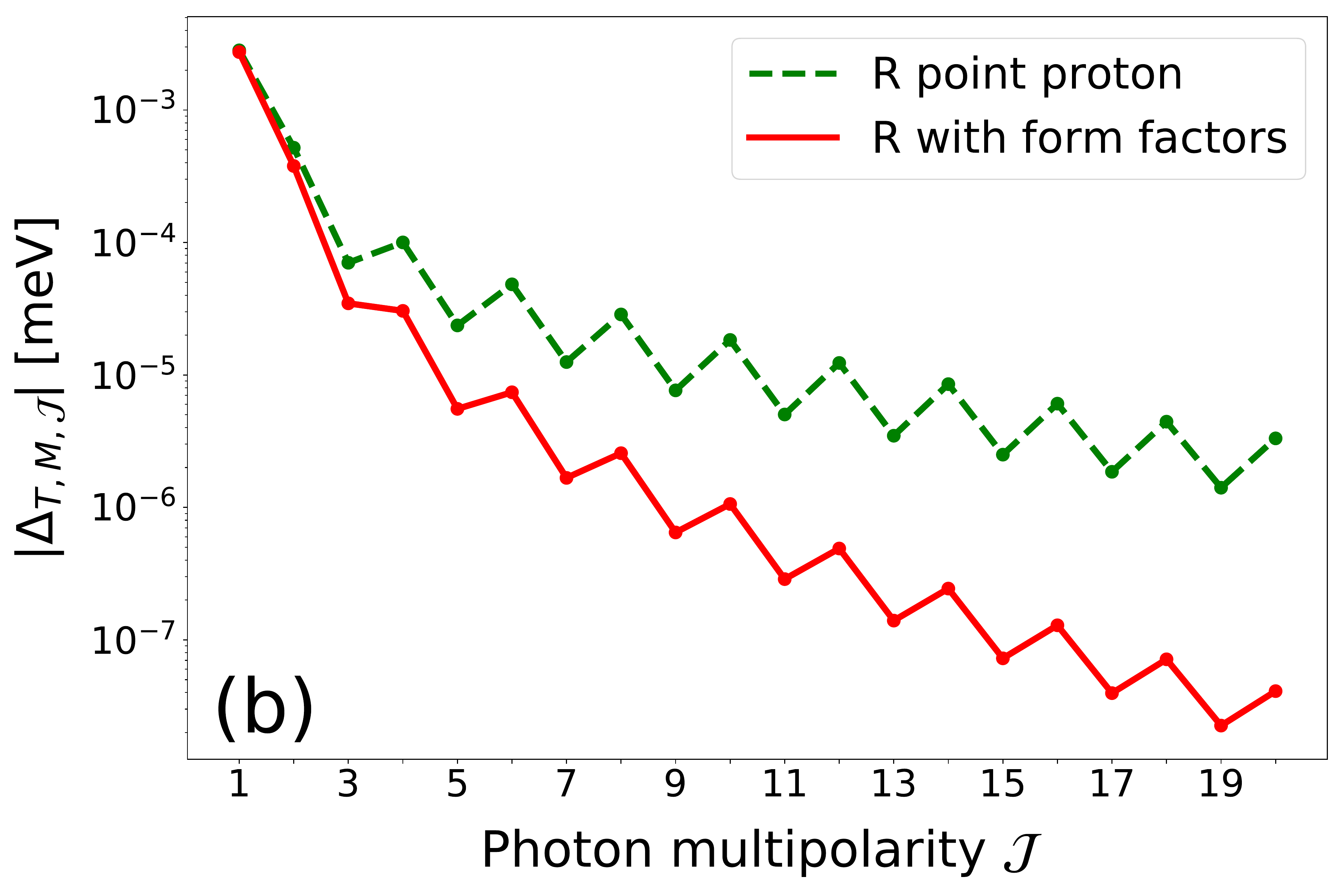} } 
 
\caption{The absolute value of the (a) transverse Siegert corrections and magnetic transverse  corrections (b) plotted  as a function of the photon multipolarity $\mathcal{J}$ for $\mu D$ in the point proton limit and with nucleon form factors.} \label{fig: convergence of transverse responses}
\end{figure}

The convergence of $\Delta_L, \Delta_{T,{\rm el}}, \Delta_{T,{\rm mag}}$ is shown in Figs.~\ref{fig: Convergence of Delta_L} and \ref{fig: convergence of transverse responses} by the decrease of the magnitudes of $\Delta_L, \Delta_{T,{\rm E},\mathcal{J}}, \Delta_{T,{\rm M},\mathcal{J}}$ with increasing photon multipoles $\mathcal{J}$.
The inclusion of nucleon form factors results in a staggering convergence pattern. This behaviour can be understood by the following argument. The Coulomb multipole tensor operator for the deuteron is
\begin{align}
C_{\mathcal{J}}(q) =& \frac{1}{2} \left[ 1+(-1)^{\mathcal{J}}\right] j_{\mathcal{J}}\left(\frac{qr}{2}\right)Y^{\mathcal{J}}(\hat{r})\notag\\
&+ \frac{1}{2} \left[\tau^3_1+(-1)^{\mathcal{J}}\tau^3_2\right] j_{\mathcal{J}}\left(\frac{qr}{2}\right)Y^{\mathcal{J}}(\hat{r}).
\end{align}
where $j_{\mathcal{J}}$ is the spherical bessel function and $\tau^{3}_{1/2}$ is the third component of the isospin for nucleons $1$ and $2$, respectively. When acting on the deuteron ground state, $(T_0=0,M_{T_0}=0,S_0=1,J_0=1,L_0=0/2)$, the reduced matrix elements for the even multipoles are isoscalar operators, while odd multipoles of this operator are isovector transitions
\begin{align}
 & \langle J;T, M_T  || C_{2\mathcal{J}}(q)  || J_0; 0,0  \rangle 
 \notag\\
  =& \langle J  || j_{2\mathcal{J}}\left(\frac{qr}{2}\right)Y^{2\mathcal{J}}(\hat{r})|| J_0  \rangle\delta_{T,0}\delta_{m_T,0},
  \\
& \langle J; T, M_T  || C_{2\mathcal{J}+1}(q)  || J_0; 0,0  \rangle 
 \notag\\
 =&   \langle J  || j_{2\mathcal{J}+1}\left(\frac{qr}{2}\right)Y^{2\mathcal{J}+1}(\hat{r})|| J_0  \rangle \delta_{T,1}\delta_{m_T,0}.
\end{align}
The above reduced matrix elements give similar numerical results accounting for the smooth convergence pattern in Fig.~\ref{fig: Convergence of Delta_L} for the point-proton limit. When nucleon form factors are introduced, the matrix elements become
\begin{align}
 &\langle J; T, M_T  || C_{2\mathcal{J}}(q)  || J_0; 0,0  \rangle  \notag\\
=&\left[ G^p_E(q^2)+G^n_E(q^2)\right] 
  \langle J  || j_{2\mathcal{J}}\left(\frac{qr}{2}\right)Y^{2\mathcal{J}}(\hat{r})|| J_0  \rangle 
  \nn
 &\times \delta_{T,0}\delta_{m_T,0}, 
 \label{eq: Coulomb isoscalar ME with FF}
\end{align}
for the even Coulomb multipoles, while the matrix elements of the odd Coulomb multipoles are
\begin{align}
& \langle J; T, M_T  || C_{2\mathcal{J}+1}(q)  || J_0; 0,0  \rangle 
\nn
=& \left[ G^p_E(q^2)-G^n_E(q^2)\right]
 \langle J  || j_{2\mathcal{J}+1}\left(\frac{qr}{2}\right)Y^{2\mathcal{J}+1}(\hat{r})|| J_0  \rangle 
\nn
&\times \delta_{T,1}\delta_{m_T,0}. 
 \label{eq: Coulomb isovector ME with FF}
\end{align}
The addition between $q$-dependent proton and neutron form factors enhances the value of $\Delta_{L,\mathcal{J}}$ with an even $\mathcal{J}$ with respect to the one with an odd $\mathcal{J}$. The enhancement becomes more pronounced at larger even values of $\mathcal{J}$, and is observed in Fig.~\ref{fig: Convergence of Delta_L} and Fig.~\ref{fig: Convergence of Delta_T_el}.

A similar pattern is observed in Fig.~\ref{fig: Convergence of Delta_T_mag}, however, in this case it is not due to the inclusion of nucleon form factors, but to the fact that for even $\mathcal{J}$ values, the deuteron transitions are dominated by isovector-spin preserving transitions while odd $\mathcal{J}$ values are dominated by isovector-spin changing operators that are suppressed relative to the former transitions. 

The final results for $\delta^A_{\rm pol}$ along with the results in the non-relativistic point proton limit $\delta^{\rm NR}_{\rm pol}$ and in the relativistic point proton limit $\delta^{\rm R}_{\rm pol}$ are given in Table \ref{table: eta comparison} in comparison to the $\eta$-expansion values.
We observe that in both the non-relativistic and relativistic cases, $\delta^A_{\rm Zem}$ and $\delta^{\rm NR/R}_{\rm pol}$ in the point nucleon limit have excellent agreement ($<0.15\%$) between the $\eta$ and $\eta$-less methods. However, when nucleon form factors are included, then $\delta^A_{\rm Zem}$ and $\delta^A_{\rm pol}$ differ by $4\%$ and $2\%$, respectively, from the $\eta$-expansion result. Although these changes appear large, when their contributions are added for $\delta^A_{\rm TPE}$, the difference reduces to $0.2\%$ between the $\eta$ and $\eta$-less results.

\begin{table}[h]
\begin{tabular}{cccc} \hline\
  &  & {$\eta$-expansion} & {$\eta$-less}  \\ \hline  \vspace{-2mm}
  &  & & \\
\multirow{3}{*}{Point proton} &$\delta^{\rm NR}_{\rm pol}$    & -1.328 & -1.326 \\
&$\delta^{A}_{\rm Zem}$    & -0.359 & -0.359 \\
&$\delta^{A}_{\rm TPE}$    & -1.687 & -1.685 \\ \hline  \vspace{-2mm}
& & & \\
\multirow{3}{*}{Point proton}  &$\delta^{\rm R}_{\rm pol}$    & -1.308 & -1.309 \\
&$\delta^{A}_{\rm Zem}$    & -0.359 & -0.359 \\
&$\delta^{A}_{\rm TPE}$    & -1.667 & -1.668 \\ \hline  \vspace{-2mm}
 & & \\
\multirow{3}{*}{R$+$FF} &$\delta^{A}_{\rm pol}$    & -1.248 & -1.269 \\ 
&$\delta^{A}_{\rm Zem}$    & -0.423 & -0.406 \\ 
&$\delta^{A}_{\rm TPE}$    & -1.671 & -1.675 \\
\vspace{-2mm}
\\ \hline\hline 
\end{tabular}
\caption{A comparison of the results from the $\eta$-formalism to the full $\eta$-less formalism in the non-relativistic point-proton, relativistic point-proton and relativistic with exact nucleon form factor (R$+$FF) calculations in units of meV, with $\delta^{A}_{\rm TPE} = \delta^{A}_{\rm pol} + \delta^{A}_{\rm Zem} $. The Coulomb correction $\delta^{(0)}_C$ = 0.262 meV from \cite{Ji_2018}, not treated here, has also been added to $\delta^{\rm NR}_{\rm pol},\delta^{\rm R}_{\rm pol},\delta^{A}_{\rm pol}$, for comparison}
\label{table: eta comparison}
\end{table}

\section{Conclusion}

In this work, we have generalized the formalism of Refs.~\cite{Rosenfelder_1983,Leidemann_1995} and made it more tractable for the study of the TPE in muonic atoms and obtained what we call the $\eta$-less expansion formalism. The calculations are done for both non-relativistic and relativistic cases. We have shown how the elastic terms implicitly included in the more general formalism can be separated from the calculations of the inelastic nuclear contributions. We applied this formalism to muonic deuterium in the non-relativistic limit and full relativistic cases with and without nucleon form factors and compared the results against those obtained from the $\eta$-expansion. Comparing the results for $\delta^{\rm NR}_{\rm pol}$ in Table \ref{table: eta comparison} amounts to a difference of only $0.1 \%$ between the $\eta$-less and $\eta$-expansion methods. This difference is from the higher order corrections included in the $\eta$-less formalism indicating that without nucleon form factors the nuclear structure calculations are reliable. The inclusion of nucleon form factors into the $\eta$-less calculations produces a discrepancy of $4\%$ and $2\%$ for the $\delta^A_{\rm Zem}$ and $\delta^A_{\rm pol}$ terms, respectively, in comparison to the $\eta$-expansion results \cite{Ji_2018}. For the latter, the form factors are approximated in a low-$q$ expansion and truncated at linear order in $q^2$. Since the $\eta$-less formalism employs exact nucleon form factors these differences indicate that the linear approximations of the nucleon form factors, used in the $\eta$-expansion~\cite{Ji_2018}, produce a higher-than-expected systematic uncertainty. However, this low-$q$ truncation uncertainty is removed when their sum $\delta^A_{\rm TPE}$ is considered. We found that the calculation of $\delta^A_{\rm TPE}$ for muonic deuterium with embedded nucleon form factors is different from the $\eta$-expansion by only about $0.2 \%$, which fully justifies the approximations made in the $\eta$-expansion for $\mu$D. This difference represents the systematic uncertainty of the $\eta$-formalism for muonic deuterium which has, for the first time, been rigorously assessed.

Combining with the Lanczos sum rules, the $\eta$-less expansion is a promising computational framework that can be applied to muonic atoms with heavier nuclei. This method can systematically compute $\delta^A_{\rm TPE}$ to higher orders and with better precision than calculations done in $\eta$ formalism. Therefore this computational framework will be an important tool for future studies of nuclear structure effects. 

\section{Acknowledgments}

The authors would like to thank Nir Nevo Dinur for insightful discussions. This work has been supported by the Cluster of Excellence ``Precision Physics, Fundamental Interactions, and Structure of Matter" (PRISMA+ EXC 2118/1) funded by the German Research Foundation (DFG) within the German Excellence Strategy (Project ID 39083149), and by the National Natural Science Foundation of China (Grant No. 11805078).

\appendix

\section{Analytical $\cancel{\pi}$EFT}
\label{Section: Pionless-EFT at N2LO }

In $\cancel{\pi}$EFT at next-to-next-to-leading order, the deuteron wave function is in the $S$-wave spin-triplet state only \cite{Friar_2013}. This allows $\delta^A_{\rm pol}$ to be calculated analytically. In this section, we show how this approximation can be used to carry out the calculation of the nuclear TPE as outlined in this work. In this formalism, the deuteron wave function is
\begin{align}
\langle \myvec{r}|N_0 \rangle &= \frac{A_s}{\sqrt{4\pi}} \frac{e^{-\kappa r}}{r},
\end{align}
where 
\begin{align}
\kappa &= \sqrt{2\mu|E_0|} /\hbar, \\
\mu &= \frac{m_p m_n}{m_p+m_n} = \frac{m_N}{2}.
\end{align}
Here, $E_0$= -2.224575(9) MeV is the deuteron binding energy, $A_s$=0.8845(8) fm$^{-1/2}$ is the asymptotic normalization constant and $\mu$ is the reduced proton and neutron mass. The summation over intermediate excited states can be performed by introducing plane waves
\begin{align}
\sum_{N \neq N_{0}} |N \rangle \langle N | &= \int \frac{d^{3}k}{(2\pi)^3} |\myvec{k} \rangle \langle \myvec{k}| - |N_{0} \rangle \langle N_{0}|,
\end{align}
where the nuclear excitation energy is
\begin{align}
\omega_k &= T_k - E_0 = \frac{k^2}{2\mu}+\frac{\kappa^2}{2\mu} .
\end{align}
The matrix elements to evaluate $\delta^A_{\rm pol}$ in $\cancel{\pi}$EFT are
\begin{align}
\langle N_0 |1 -e^{i \frac{1}{2}\myvec{q}\cdot \myvec{r} }| \myvec{k} \rangle 
 =& \sqrt{4\pi} A_s 
  \left[ \frac{1}{\kappa^2 + |\myvec{k}+\frac{1}{2}\myvec{q}|^2} - \frac{1}{\kappa^2 +k^2}\right],
\end{align}

\begin{align}
\langle N_0 |1 -e^{i \frac{1}{2}\myvec{q}\cdot \myvec{r} }|  N_0 \rangle &= A^2_s \left[  \frac{1}{2\kappa} - \frac{2}{q}\tan^{-1}\left( \frac{q}{4\kappa} \right)\right], \\
\langle N_0 |e^{i \frac{1}{2}\myvec{q}\cdot \myvec{r} }|  N_0 \rangle &= \frac{2 A^2_s }{q}\tan^{-1}\left( \frac{q}{4\kappa} \right).
\end{align}
These expressions are related to the $\eta$-less method in section \ref{Section: covariant etaless expansion}, through a multipole decomposition of the matrix elements. This decomposition is carried out by and introducing the $g_\mathcal{J}(k,q)$ functions
\begin{align}
\frac{1}{\kappa^2+|\myvec{k}+\frac{1}{2}\myvec{q}|^2} - \frac{1}{\kappa^2+k^2} &= \sum_{\mathcal{J}=0} g_{\mathcal{J}}(k,q)P_{\mathcal{J}}(x), 
\end{align}
with
\begin{align}
 g_{\mathcal{J}}(k,q) =&
  \frac{2\mathcal{J}+1}{2}\int\limits_{-1}^1 dx \ P_\mathcal{J}(x)
 \nn 
 &\times
 \left[\frac{1}{\kappa^2+k^2+\frac{1}{4}q^2+ k q x} - \frac{1}{\kappa^2+k^2}  \right].
\end{align}
Integrating out the angle $\hat{k}$, we have
\begin{align}
\int d\hat{k} \ |\langle N_0 |1 -e^{i \frac{1}{2}\myvec{q}\cdot \myvec{k} }| \myvec{k} \rangle|^2 &= 4 \pi A^2_s  \sum_{\mathcal{J}=0} \frac{4\pi}{2\mathcal{J}+1} g^2_{\mathcal{J}}(k,q). 
\end{align}
Combining all of these expressions, the leading order non-relativistic point nucleon corrections are given by
\begin{align}
\delta^{\rm NR}_{\rm pol} =&  -8\alpha^2|\phi_\mu(0)|^2 \int\limits_{0}^{\infty} \frac{dq}{q^2} 
\nn
&\times \left\{ \sum_{\mathcal{J}=0}^{\infty} \frac{4\pi A^2_s}{(2\mathcal{J}+1)}\int\limits_{0}^{\infty} \frac{k^2 dk}{(2\pi)^3}   K_{\rm NR}(q,\omega_k) g^2_{\mathcal{J}}(k,q) \right. 
\nn
 & \left. -  K_{\rm NR}(q,0)A^4_{s}\left[\frac{1}{2\kappa}-\frac{2}{q}\tan^{-1}\left( \frac{q}{4\kappa} \right) \right]^2 \right\}. 
 \label{eq: Pionless-EFT, non-relativistic point nucleon result}
\end{align}

\section{Finite nucleon size corrections in the $\eta$-less expansion}\label{Section: finite size corrections}

In this appendix, we show that when nucleon form factors are included, $\delta^{A}_{\rm pol}$ becomes 
\begin{equation}
\delta^A_{\rm pol} = \Delta^{\rm inel}_{\omega}+ \delta^{(1)}_{Z3}+\delta^{(1)}_{Z1}+\delta^{(1)}_{R3}+\delta^{(1)}_{R1}
\end{equation}
where $\delta^{(1)}_{Z3}$ is the third-Zemach moment of the nucleus in Eq.~\eqref{eq: Definition of Z3}, $\delta^{(1)}_{R3}$ is the term defined in Eq.~\eqref{eq: Definition of R3} and the terms $\delta^{(1)}_{Z1}$, $\delta^{(1)}_{R1}$ are the finite nucleon size corrections to those terms
\begin{align}
\delta^{(1)}_{R1} =& -\frac{\pi}{3}m_r(Z\alpha)^2|\phi_\mu(0)|^2  
\nn
&\times  \left[\langle Y_{pp}^3\rangle_{(2)}+\langle Y_{nn}^3\rangle_{(2)}+2\langle Y_{np}^3\rangle_{(2)} \right], 
\\
\delta^{(1)}_{Z1} =& \frac{\pi}{3}m_r(Z\alpha)^2|\phi_\mu(0)|^2 
\nn
&\times  \left[\langle Z_{pp}^3\rangle_{(2)}+\langle Z_{nn}^3\rangle_{(2)}+2\langle Z_{np}^3\rangle_{(2)} \right].
\end{align}
The terms $\langle Y_{ab}^3\rangle_{(2)}$ and $\langle Z_{ab}^3\rangle_{(2)}$ with $(a,b) = (p,p),(n,p),(n,n)$ denote the following integrals
\begin{align}
\langle Y_{pp}^3\rangle_{(2)} =& \frac{48}{\pi}  \int\limits_0^\infty  \frac{dq}{q^4} \left[ G^p_E(q^2)^2-1 \right] 
\nn
&\times 
\left[ \langle N_0| \tilde{\rho}^p(q)\tilde{\rho}^{p,\dagger}(q) | N_0 \rangle -1 \right],
\end{align}
\begin{align}
\langle Y_{np}^3\rangle_{(2)} =& \frac{48}{\pi} \int\limits_0^\infty  \frac{dq}{q^4} G^n_E(q^2)
\nn 
&\times
\left[ \langle N_0| \tilde{\rho}^p(q)\tilde{\rho}^{n,\dagger}(q) | N_0 \rangle G^p_E(q^2) -1 \right],
\end{align}
\begin{align}
\langle Y_{nn}^3\rangle_{(2)} =& \frac{48}{\pi} \int\limits_0^\infty  \frac{dq}{q^4} G^n_E(q^2)^2
\left[ \langle N_0| \tilde{\rho}^n(q)\tilde{\rho}^{n,\dagger}(q) | N_0 \rangle -1 \right],
\end{align}
\begin{align}
\langle Z_{pp}^3\rangle_{(2)} =& \frac{48}{\pi} \int\limits_0^\infty  \frac{dq}{q^4} \left[G^p_E(q^2)^2-1\right]
\nn 
&\times
\left[ |\langle N_0| \tilde{\rho}^p(q)| N_0 \rangle|^2 -1 \right], 
\end{align}
\begin{align}
\langle Z_{np}^3\rangle_{(2)} =& \frac{48}{\pi} \int\limits_0^\infty  \frac{dq}{q^4} G^n_E(q^2)
\nn
\times&
 \left[ \langle N_0| \tilde{\rho}^p(q) | N_0 \rangle\langle N_0| \tilde{\rho}^{n,\dagger}(q) | N_0 \rangle G^p_E(q^2) -1 \right],
\end{align}
\begin{align}
\langle Z_{nn}^3\rangle_{(2)} = \frac{48}{\pi} \int\limits_0^\infty  \frac{dq}{q^4} G^n_E(q^2)^2
 \left[ |\langle N_0| \tilde{\rho}^n(q) | N_0 \rangle|^2 -1 \right].
\end{align}

The above expressions correspond to the same expressions $\delta^{(1)}_{Z1},\delta^{(1)}_{R1}$ as in Ref.~\cite{Ji_2018} in the low-$q$ limit of the nucleon form factors where $G^p_E(q^2) \approx 1 - 2 q^2/\beta^2$ and $G^n_E(q^2) \approx \lambda q^2$.
\begin{center}
\begin{table}[h]
\begin{tabular}{ccc} \hline\
    & {$\eta$-expansion } & {$\eta$-less }  \\ \hline  \vspace{-2mm}
    & & \\
$\delta^{(1)}_{Z3}$    & 0.359 & 0.359  \\
$\delta^{(1)}_{Z1}$    & 0.064 & 0.047 \\
$\delta^{(1)}_{R1}$    & 0.017 & 0.016 \\ \hline 
\end{tabular}
\caption{A comparison between the nucleon size corrections obtained in the $\eta$-less formalism with the approximated nucleon form factors to the $\eta$-less expansion results with exact nucleon form factors in units of meV.}
\label{table: eta comparison Elastic terms}
\end{table}
\end{center}

\section{Multipole expansion}\label{section: multipole expansion}

To compute the expressions in Eqs.~\eqref{eq: Longitudinal Correction}, \eqref{eq: Transverse Electric Correction}, and \eqref{eq: Transverse Magnetic Correction}, the charge density $\tilde{\rho}(q)$ and current density $\hat{e}^\dagger_{\lambda} \cdot \myvec{\tilde{J}}(q)$ are expanded in plane waves as
\begin{align}
\tilde{\rho}(q) =& \sum\limits_{\mathcal{J} \geq 0 } \sqrt{4\pi(2\mathcal{J}+1)} i^{\mathcal{J}}C_{\mathcal{J}}(q),  \\
\hat{e}^\dagger_{\lambda} \cdot\myvec{\tilde{J}}(q) =& -\sum\limits_{\mathcal{J} \geq 1 } \sqrt{2\pi(2\mathcal{J}+1)} i^{\mathcal{J}} 
\nn
&\times
 \left[ T^{\rm E}_{\mathcal{J}-\lambda}(q) +\lambda  T^{\rm M}_{\mathcal{J}-\lambda}(q)  \right].  
\end{align}
The operators $C_{\mathcal{J}}(q)$ are the Coulomb multipole operators
\begin{equation}
C_{\mathcal{J}}(q) = \int d^3x \ j_\mathcal{J}(qx)Y^0_\mathcal{J}(\hat{x})\rho(\myvec{x}),
\end{equation}
and $T^{\rm E/ \rm M}_{\mathcal{J} \lambda}(q)$ are the transverse electric/magnetic tensor operators, respectively, with photon multipolarity $\mathcal{J}$ as in Ref.~\cite{Walecka_2004}
\begin{align}
 T^{\rm E}_{\mathcal{J} \lambda}(q) =& \frac{1}{q} \int d^3 x \ \left[\myvec{\nabla} \times j_{\mathcal{J}}(qx)\myvec{Y}^{\lambda}_{\mathcal{J}\mathcal{J} 1}(\myvec{x}) \right]\cdot \myvec{J}_c(\myvec{x}), \label{eq: Electric Tensor Definition}\\
 T^{\rm M}_{\mathcal{J} \lambda}(q) =& \int d^3 x \ \left\{ j_{\mathcal{J}}(qx)\myvec{Y}^{\lambda}_{\mathcal{J}\mathcal{J} 1}(\hat{x})\cdot \myvec{J}_c(\myvec{x})  \right.\notag\\
&\left.+ \left[\myvec{\nabla} \times j_{\mathcal{J}}(qx)\myvec{Y}^{\lambda}_{\mathcal{J}\mathcal{J} 1}(\myvec{x}) \right]\cdot \myvec{J}_s(\myvec{x}) \right\} . \label{eq: Magnetic Tensor Definition} 
\end{align}
The functions $\myvec{Y}^\lambda_{\mathcal{J}\mathcal{J}1}(\myvec{x})$ are the spherical vector harmonics, while $\myvec{J}_s(\myvec{x})$ is the nuclear spin current and $\myvec{J}_c(\myvec{x})$ is the convection current \cite{Ericson_1988}. Using this expansion, the longitudinal $S_L$ and transverse $S_T$ response functions in Section \ref{Section: covariant etaless expansion} are expanded as sums of multipole functions with photon multipolarities $\mathcal{J}$
\begin{align}
S_{L}(q,\omega) &= \sum\limits_{\mathcal{J}=0}^\infty S_{L,\mathcal{J}}(q,\omega),\label{eq: longitudinal response} \\
S^{\rm E/M}_{T}(q,\omega) &= \sum\limits_{\mathcal{J}=1}^\infty S^{\rm E/M}_{T,\mathcal{J}}(q,\omega)\label{eq: transverse response}.
\end{align}
The transverse response function is separated into the sum of electric and magnetic terms
\begin{align}
S_{T}(q,\omega) &= S^{\rm E}_{T}(q,\omega)+ S^{\rm M}_{T}(q,\omega).  
\end{align}  
The expressions for each multipole function are 
\begin{align}
S_{L,\mathcal{J}}(q,\omega) =& \frac{4\pi}{2J_0+1}\sum_{N\neq N_0}|\langle N ||C_{\mathcal{J}}(q)|| N_0 \rangle|^2
\nn
&\times \delta(\omega_N - \omega),\label{eq: longitudinal response definition}
\end{align}
\begin{align}
S^{\rm E/M}_{T,\mathcal{J}}(q,\omega) =& \frac{4\pi}{2J_0+1} \sum_{N \neq N_0} |\langle N || T^{\rm E/M}_\mathcal{J}(q) || N_0 \rangle|^2 
\nn
&\times 
\delta(\omega_N - \omega).\label{eq: Electric/Magnetic transverse response definition}
\end{align}
 In this work, we apply the Siegert theorem to relate the electric transverse multipole functions $S^{\rm E}_{T,\mathcal{J}}$ to the longitudinal ones through
\begin{equation}
S^{\rm E}_{T,\mathcal{J}}(q,\omega) =  \frac{\omega^2}{q^2}\left[ \frac{\mathcal{J}+1}{\mathcal{J}} \right]S_{L,\mathcal{J}}(q,\omega)
+\delta S_{T,\mathcal{J}}(q,\omega),
\end{equation}
where $\delta S_{T,\mathcal{J}}(q,\omega)$ is the correction to the Siegert approximation. In Refs.~\cite{Bacca_2007,Bacca_2014} this correction has been shown to be negligible in the low momentum region that dominate the integrals of Eq.~\eqref{eq: Total two photon exchange correction} and will not be included in this analysis.

In the low-$q$ limit, the Siegert approximated transverse electric response function reduces to only the dipole part with $\mathcal{J}=1$:
\begin{align}
S^{\rm E}_{T,\mathcal{J}}(0,\omega) =& S^{\rm E}_{T,1}(0,\omega) \delta_{\mathcal{J},1}
= \frac{2\omega^2}{9} 
R_{D}(\omega) \delta_{\mathcal{J},1},
\label{eq: Transverse Siegert Limit}
\end{align}
where $R_{D}(\omega)$ is the electric dipole function of the nucleus,
\begin{align}
R_{D}(\omega) =& \frac{4\pi}{2J_0+1}\sum_{N \neq N_0} 
|\langle N ||  \sum_{i=1}^A \hat{p}_{i} r_i Y^{1}(\hat{r}_i) || N_0 \rangle|^2
\nn
&\times \delta(\omega_N - \omega),
\end{align}
the operators $\hat{p}_i$ are the proton charge projection operators. Because Eq.~\eqref{eq: Transverse Siegert Limit} is non-zero at $\mathcal{J}=1$, $q=0$ and the transverse Kernel is singular at low-$q$, the dipole contributions in Eq.~\eqref{eq: Transverse Electric Correction} requires regularization. The seagull term in Fig.~\ref{fig:all two-photon exchange diagrams} removes the singularity of the transverse Kernel for this multipole. Higher electric multipole response functions vanish in the $q \rightarrow 0$ limit and do not require the Seagull term. Similarly, the transverse magnetic response function $S^{\rm M}_{T}(q,\omega)$ vanishes in the low-$q$ limit for all multipoles and does not require the regularization term.\\

\bibliography{mybibfile}

\end{document}